\documentclass[
 amsmath,amssymb,
 aps,
prstab,
]{revtex4-2}

\usepackage{graphicx}
\usepackage{dcolumn}
\usepackage{bm}

\usepackage{xcolor}

\begin{document}
\preprint{APS/123-QED}

\title{Beam-excited resonant modes in RF cavities}

\author{Yanxu Wang}
\affiliation{%
 Shanghai Institute of Applied Physics, Shanghai 201800, China,
 University of Chinese Academy of Sciences, Beijing 101408, China}%

\author{Demin Zhou}
\email{dmzhou@post.kek.jp}
\affiliation{%
 High Energy Accelerator Research Organization (KEK),  1-1 Oho, Tsukuba 305-0801, Ibaraki, Japan}
 
\author{Ji Li}
\affiliation{Zhangjiang Laboratory, Shanghai 201210, China}
 
\author{Weijie Fan}
\email{fanwj@sari.ac.cn}
\author{Qinglei Zhang}
\email{zhangql@sari.ac.cn}

\author{Chao Feng}
\author{Zhentang Zhao}
\affiliation{Shanghai Advanced Research Institute, Shanghai 201210, China}%

\date{\today}

\begin{abstract}
Beam-excited resonant modes in RF cavities are important sources of beam-coupling impedance and coupled-bunch instabilities in high-current storage rings. We develop a unified framework for longitudinal and transverse resonant impedances based on Maxwell's equations and generalized cavity-voltage definitions, and derive analytical expressions for impedances obtained from finite-length truncated wakefields. The formulation enables the resonant frequencies, normalized longitudinal and transverse shunt impedances, and, when sufficiently constrained, the quality factors to be extracted from practical wakefield simulations without requiring fully converged long-range wakes. The method is validated with an axisymmetric pillbox cavity through comparison with analytical results and eigenmode calculations. It is then applied to the RF cavity of the Storage-Ring-based Coherent Light Source (SRCLS), where the extracted HOM parameters are used to reconstruct total impedance spectra, evaluate coupled-bunch instability thresholds, and guide cavity-geometry optimization. The results demonstrate an efficient connection between wakefield analysis, eigenmode characterization, and beam-stability evaluation for practical RF-cavity designs.

\end{abstract}

\maketitle

\section{\label{sec:introduction}INTRODUCTION}

In particle accelerators, interactions between the beam and its surrounding environment play an important role in determining beam quality and stability. These interactions are commonly described in terms of wakefields and impedances, for which a well-established theoretical framework exists~\cite{heifets1991coupling, chao1993physics, zotter1998impedances}. Among the various accelerator components, RF cavities are particularly important sources of narrowband impedance. In addition to the fundamental accelerating mode, the passage of a charged beam can excite higher-order resonant modes. If these modes are not sufficiently damped, the associated long-range wakefields may drive coupled-bunch instabilities~\cite{chao1993physics} and thereby limit the achievable beam current and machine performance. A reliable characterization of RF-cavity resonant modes is therefore an essential part of accelerator design.

A resonant mode of an RF cavity is usually characterized by its resonant wavenumber ($k_r$), shunt impedance ($R_s$), and quality factor ($Q$). These parameters provide the basic input for evaluating coupled-bunch instabilities~\cite{MarhauserF2001HOM}, beam-induced heating~\cite{day2012beam}, and possible mitigation measures~\cite{KangHS2003Suppression}. For narrowband cavity impedances, the standard RLC resonator model is widely used~\cite{zotter1998impedances, wolski2014beam, lee2018accelerator}. In the loss-free limit of a closed cavity, corresponding to $Q\rightarrow\infty$, the resonant response is determined by the mode frequency and the normalized shunt impedance $R_s/Q$~\cite{weiland1980wake}. A general formulation based on normal-mode expansion was developed in Ref.~\cite{bane1984wake}, providing a theoretical foundation for relating cavity eigenmodes to beam-coupling impedances.

In practical applications, the mode parameters can be obtained either from frequency-domain eigenmode calculations or from time-domain wakefield simulations. Source-free eigenmode solvers, such as the CST Eigenmode Solver~\cite{CST} and GdfidL~\cite{gdfidl}, directly compute the resonant frequencies and field distributions of cavity modes; an example of such an approach is given in Ref.~\cite{arsenyev2019method}. Alternatively, wakefields excited by a driving beam can be computed using these codes or dedicated wakefield solvers such as ECHO3D~\cite{ECHO3D}, and the mode parameters can then be extracted from the resulting wakefields or impedances. These two approaches are complementary: eigenmode calculations provide direct access to source-free cavity modes, whereas wakefield simulations describe the beam-excited response of the structure.

Both approaches, however, can become challenging for high-$Q$ resonant modes. In time-domain simulations, the wakefield must be computed over a distance much longer than the modal damping length in order to obtain a fully converged impedance spectrum. For high-$Q$ modes in RF cavities, this distance may extend to hundreds of meters or even kilometers~\cite{NgKY2010explicit}, leading to prohibitive computational costs, especially for short bunches that require fine spatial discretization. 
Frequency-domain eigenmode solvers avoid such long-distance wakefield tracking. However, realistic RF cavities containing beam pipes, input couplers, waveguides, HOM absorbers, and other auxiliary structures are generally three-dimensional (3D) and lack exact symmetry. The resulting dense spectrum of closely spaced modes complicates not only the identification of beam-relevant resonances, but also the interpretation of their longitudinal and transverse beam-coupling impedances. In particular, existing formulations often infer transverse impedances from near-axis expansions of the longitudinal shunt impedance, an approach whose applicability becomes less evident in strongly asymmetric 3D structures.
In such cases, identifying the beam-relevant modes and consistently relating eigenmode results to wakefield-based impedances can be nontrivial.

These considerations motivate a systematic formulation that connects the resonator description, finite-length wakefield calculations, and source-free eigenmode results for beam-excited cavity modes. The main purpose of this work is to develop a practical extraction method that analytically accounts for wakefield truncation, enabling beam-relevant modal parameters to be obtained from finite-length wakefield simulations and compared consistently with eigenmode calculations. This approach is particularly useful for high-$Q$ modes, for which direct computation of fully converged impedance spectra can be prohibitively expensive.

The present study is motivated by the requirements of the Storage-Ring-based Coherent Light Source (SRCLS)~\cite{feng2017storage,li2020extremely, jiang2022synchrotron, LuYujie2025Lattice}, proposed by the Shanghai Synchrotron Radiation Facility (SSRF) tea to realize the angular-dispersion-induced microbunching (ADM) scheme~\cite{feng2017storage} in storage ring. The facility uses a storage ring, with a damping time of about 1.4 ms, to provide reusable high-current beams to a bypass line for coherent-radiation generation. The beam current of the ring can reach 1 A, and the natural bunch length is about 2.9 mm. Under such high-current operating conditions, the RF cavities must be carefully designed and their resonant modes must be characterized to avoid potential coupled-bunch instabilities.

In this paper, we present a systematic study of beam-excited resonant modes in RF cavities. Following the formalism of Ref.~\cite{bane1984wake}, Sec.~\ref{sec:theory} develops the impedance theory for RF-cavity resonant modes and establishes its connection with practical wakefield simulations and eigenmode calculations. In particular, the effect of finite wakefield length on the extracted impedance spectrum is analyzed, providing a basis for determining resonant-mode parameters from truncated wakefield data. The theory is benchmarked in Sec.~\ref{sec:validation} using a simple pillbox-like cavity, for which the numerical results can be compared with analytical solutions. In Sec.~\ref{sec:application}, the impedances of the SRCLS RF cavities are studied using both wakefield and eigenmode calculations. Based on these results, cavity-geometry optimization for suppressing resonant modes is investigated, and the extracted modal parameters are used to evaluate coupled-bunch-instability thresholds for potentially dangerous HOMs. Finally, the main conclusions are summarized in Sec.~\ref{sec:summary}.

\section{\label{sec:theory}Theory}

As discussed in Sec.~\ref{sec:introduction}, a practical analysis of RF-cavity resonant modes requires a consistent connection among the modal description, wakefield calculations, and impedance parameters. Such a framework, based on Maxwell's equations and the definition of the wake function, was developed in Ref.~\cite{bane1984wake}. In this section, we extend the formulation to include field damping and derive the wake functions and corresponding impedance expressions for the resonant modes in 3D asymmetric cavity structures, including their dependence on the transverse positions of the source and test particles. The relation between longitudinal and transverse wake functions, and equivalently between their impedances, is then examined. We also clarify how the quantities defined in impedance theory are related to those obtained from eigenmode calculations. Based on these results, we formulate two methods for extracting the main resonant-mode parameters: one from impedances computed using finite-length, or truncated, wake potentials, which is particularly useful when the wake distance is limited by computational resources; and the other from eigenmode fields computed using eigenmode solvers.

\subsection{\label{sec:ResonantModes} Impedance models for resonant modes}

Suppose a 3D cavity structure supports an eigenmode 
$(\mathbf E_0(\mathbf r), \mathbf B_0(\mathbf r))$ 
with eigenfrequency $\omega_r$. A point charge $q_0$ moving along the trajectory $\mathbf r_0 = (x_0,y_0,z_0=ct)$ excites this specific eigenmode and has charge density
\begin{equation}
    \rho(\mathbf r,t)
    = q_0 \delta(x-x_0)\delta(y-y_0)\delta(z-ct),
\end{equation}
and current density
\begin{equation}
    \mathbf J(\mathbf r,t)
    = \rho \mathbf v
    = q_0 c\,\delta(x-x_0)\delta(y-y_0)\delta(z-ct)\,\hat{\mathbf z}
\end{equation}
with $\mathbf r =(x,y,z)$. Here, we adopt the ultra-relativistic limit, assuming that the point charge travels at the speed of light.

Assume that only this single mode is excited and write the fields as
\begin{equation}
    \mathbf E(\mathbf r,t)
    =
    a(t)\,\mathbf E_0(\mathbf r),
\end{equation}
and
\begin{equation}
    \mathbf B(\mathbf r,t)
    =
    a(t)\,\mathbf B_0(\mathbf r),
\end{equation}
where $a(t)$ is a scalar amplitude. We normalize the eigenmode according to the stored energy~\cite{wolski2014beam}
\begin{equation}
    U_0 = \frac{1}{2} \int_V 
     \left( \epsilon_0\mathbf E_0^2 +\frac{1}{\mu_0}\mathbf B_0^2 \right) dV,
\end{equation}
which corresponds to the maximum stored energy when the complex modal amplitude is unity. Projecting Maxwell's equations with source $\mathbf J$ onto the eigenmode yields the driven oscillator equation
\begin{equation}
\ddot a(t) + 2 \gamma \dot a(t) +\omega_r^2 a(t)
=
-\frac{1}{U_0}
\frac{d}{dt}
\int_V
\mathbf J(\mathbf r,t)\cdot\mathbf E_0(\mathbf r)\, dV,
\label{eq:Dam-fluctuation}
\end{equation}
where $\gamma\equiv \omega_r/(2Q)$ is the damping rate, and $Q=\omega_rU/P_{\rm loss}$ is the quality factor of the mode, with
$P_{\rm loss}=-dU/dt$ denoting the time-averaged power loss. With this
definition, the stored energy decays as
$U(t)=U_0 e^{-2\gamma t}$~\cite{lee2018accelerator}. For $t>L_c/c$, where $L_c$ is the cavity length in the $z$ direction, the above equation can be solved to obtain
\begin{equation}
a(t)
=
-\frac{q_0 c}{ U_0}
\int_{0}^{L_c/c}
F_c'(c(t-\tau))
E_{0z}(x_0,y_0,c\tau)
\, d\tau
\end{equation}
with $E_{0z}$ the longitudinal component of $\mathbf E_0$, and $F_c'(x)$ defined by
\begin{equation}
    F_c'(x)=e^{-\alpha x}
    \left[
    \cos\!\bigl(k_d x\bigr)
    -\frac{\alpha}{k_d}\sin\!\bigl(k_d x \bigr)
    \right],
\end{equation}
where $k_d=\sqrt{k_r^2-\alpha^2}$ with $k_r=\omega_r/c$ and $\alpha=\gamma/c$. We have also assumed zero initial excitation:
\begin{equation}
    a(-\infty)=\dot a(-\infty)=0.
\end{equation}

Consider a test particle trailing the source charge $q_0$ by a distance $s$. The longitudinal wake function is then defined as
\begin{equation}
    W_\parallel(s; \mathbf{r}_{0\perp}, \mathbf{r}_{1\perp})=-\frac{1}{q_0} \int_0^{L_c}
    dz E_z\left(x_1,y_1,z;\frac{z+s}{c}\right),
    \label{eq:Wake_function}
\end{equation}
where 
\begin{equation}
    E_z\left(x_1,y_1,z;t\right)=
    a\left(t\right) E_{0z}(x_1,y_1,z)
\end{equation}
which is obtained from the previous formulations. $\mathbf{r}_{0\perp}=(x_0,y_0)$ and $\mathbf{r}_{1\perp}=(x_1,y_1)$ denote the transverse coordinates of the source and test particles, respectively. The integration limits in
Eq.~\eqref{eq:Wake_function} are chosen such that $z=L_c/2$ corresponds to the center of the cavity, while at $z=0$ and $z=L_c$ the eigenmode fields satisfy $(\mathbf E_0(\mathbf r),\mathbf B_0(\mathbf r))\rightarrow(\mathbf{0},\mathbf{0})$. If this condition is not fulfilled, the cavity length $L_c$ should be extended accordingly.

In the high-$Q$ limit, and for sufficiently small transverse offsets of the source and test particles, the damping of the modal field during beam passage can be neglected and only the leading-order transverse dependence needs to be retained. Under these assumptions, Eq.~\eqref{eq:Wake_function} reduces to
\begin{equation}
W_{\parallel}(s; \mathbf{r}_{0\perp}, \mathbf{r}_{1\perp}) = \frac{ \omega_r}{Q} R_{\parallel}(\mathbf{r}_{0\perp}, \mathbf{r}_{1\perp}) F_c'\left( s \right),
\label{eq:WakeFunctionLongitudinal}
\end{equation}
where $R_\parallel$ has the unit of impedance and is given by
\begin{equation}
    R_\parallel (\mathbf{r}_{0\perp}, \mathbf{r}_{1\perp})
    =\frac{Q\mathcal{V}^+(\mathbf{r}_{0\perp}, \mathbf{r}_{1\perp})}{\omega_rU_0}.
    \label{eq:RLgeneral}
\end{equation}
Here, $\mathcal{V}^+$ is defined by Eq.~\eqref{eq:effectiveVoltage}. The exact solution of Eq.~\eqref{eq:Wake_function} and the approximations leading to Eq.~\eqref{eq:WakeFunctionLongitudinal} are presented in Appendix~\ref{sec:WakeFunctionApprox}. The quantity $R_\parallel$ is analogous to the longitudinal shunt impedance used in RF technology (e.g., see~\cite{padamsee2008rf}), but is generalized here to depend on the transverse offsets of both the source and test particles. The quantity $R_\parallel/Q$, on the other hand, corresponds to normalized shunt impedance. It is primarily determined by the eigenmode field pattern and cavity geometry.

Fourier transforming Eq.~\eqref{eq:WakeFunctionLongitudinal} yields the well-known resonator impedance model of
\begin{equation}
    Z_{\parallel}(k; \mathbf r_{0\perp},\mathbf r_{1\perp})=
    R_{\parallel}(\mathbf r_{0\perp},\mathbf r_{1\perp}) G_r(k;Q,k_r),
\label{eq:longitudinal impedance}
\end{equation}
with wave number $k=\omega/c$ and the resonant response factor
\begin{equation}
G_r(k;Q,k_r)
\equiv
\frac{1}{1+iQ\left( \frac{k_r}{k}-\frac{k}{k_r} \right)} .
\label{eq:resonant_response_factor}
\end{equation}

The transverse wake function is given by
\begin{equation}
\mathbf{W}_{\perp}(s; \mathbf{r}_{0\perp}, \mathbf{r}_{1\perp}) \equiv -\frac{1}{q_0} \int_{0}^{L_c} \left( \mathbf{E} + \mathbf{v} \times \mathbf{B} \right) dz,
\end{equation}
where $\mathbf{W}_{\perp}=(W_x,W_y)$. For convenience of discussion, we only consider the horizontal wake function
\begin{equation}
\begin{aligned}
W_x(s; \mathbf{r}_{0\perp}, \mathbf{r}_{1\perp}) = & -\frac{1}{q_0} \int_{0}^{L_c} \Bigg[  E_x\left(x_1, y_1, z; \frac{z+s}{c}\right) \\
    &- c B_y\left(x_1, y_1, z; \frac{z+s}{c}\right) \Bigg] dz,
\end{aligned}
\end{equation}
with
\begin{equation}
\begin{aligned}
    E_x\left(x_1,y_1,z;t\right) &= a\left(t\right) E_{0x}(x_1,y_1,z) \\
    B_y\left(x_1,y_1,z;t\right) &= a\left(t\right) B_{0y}(x_1,y_1,z).
\end{aligned}
\end{equation}
Retaining the leading-order terms in the high-$Q$ and small-offset limits, the horizontal wake function can be obtained as
\begin{equation}
W_{x}(s; \mathbf{r}_{0\perp}, \mathbf{r}_{1\perp}) = \frac{\omega_r}{Q} R_{x}(\mathbf{r}_{0\perp}, \mathbf{r}_{1\perp}) F_c\left( s \right),
\label{eq:tra_wake_function_x}
\end{equation}
with
\begin{equation}
    F_c(x)=\frac{1}{k_d}e^{-\alpha x}\sin\left( k_d x \right),
\end{equation}
and
\begin{equation}
R_{x}(\mathbf{r}_{0\perp}, \mathbf{r}_{1\perp}) 
= \frac{\partial R_{\parallel}(\mathbf{r}_{0\perp}, \mathbf{r}_{1\perp})}{\partial x_1}.
\end{equation}
The quantity $R_x$ is analogous to the transverse shunt impedance of the cavity, but is generalized here to depend on the transverse offsets of both the source and test particles. The vertical wake function can be formulated similarly. The resulting transverse and longitudinal wake functions satisfy the Panofsky-Wenzel theorem~\cite{panofsky1956some}:
\begin{equation}
\displaystyle
\nabla_{\perp,1} W_{\parallel}(s;\mathbf{r}_{0\perp}, \mathbf{r}_{1\perp}) = \frac{\partial \mathbf W_{\perp}(s;\mathbf{r}_{0\perp}, \mathbf{r}_{1\perp})}{\partial s},
\end{equation}
which leads to the relation of impedances:
\begin{equation}
    \nabla_{\perp,1} Z_\parallel(k;\mathbf{r}_{0\perp}, \mathbf{r}_{1\perp})
    =
     k \mathbf Z_\perp(k;\mathbf{r}_{0\perp},\mathbf{r}_{1\perp}),
    \label{eq:P-W-theorem-impedance}
\end{equation}
Here, $\mathbf Z_\perp=(Z_x,Z_y)$ is the transverse impedance vector, and $\nabla_{\perp,1}=(\partial/\partial x_1,\partial/\partial y_1)$ acts on the test-particle coordinates. The horizontal impedance can be explicitly expressed as
\begin{equation}
Z_{x}(k; \mathbf{r}_{0\perp}, \mathbf{r}_{1\perp}) 
 = \frac{1}{k} R_{x}(\mathbf r_{0\perp},\mathbf r_{1\perp}) G_r(k;Q,k_r).
\end{equation}
A similar formulation can be derived for the vertical direction, but is omitted here for brevity.

In practical instability analyses, it is sufficient to retain only the low-order terms in a transverse Taylor expansion of the impedance around the reference orbit. The transverse impedances can therefore be written as~\cite{heifets1997generalized}
\begin{equation}
    Z_x(k; \mathbf{r}_{0\perp}, \mathbf{r}_{1\perp})
    =
    Z_{xm}(k)
    + Z_{xd}(k) x_0
    + Z_{xq}(k) x_1
    + \cdots,
\label{eq:Timpedance_all}
\end{equation}
\begin{equation}
    Z_y(k; \mathbf{r}_{0\perp}, \mathbf{r}_{1\perp})
    =
    Z_{ym}(k)
    + Z_{yd}(k) y_0
    + Z_{yq}(k) y_1
    + \cdots,
\end{equation}
where $Z_{um}$, $Z_{ud}$, and $Z_{uq}$ ($u=x,y$) denote the monopolar, dipolar (driving), and quadrupolar (detuning) components, respectively. These terms have clear physical interpretations: the monopolar impedance causes a static tilt of the bunch, the dipolar impedance drives coherent beam motion, while the quadrupolar term contributes to tune shifts and detuning. The above expansion assumes that the closed orbit at the impedance location is zero; otherwise, the expansion must be performed about the actual orbit (see~\cite{arsenyev2019method} for a related discussion). The Fourier transforms of $Z_{um}(k)$, $Z_{ud}(k)$, and $Z_{uq}(k)$ define the corresponding monopolar, dipolar, and quadrupolar wake functions, denoted by $W_{um}$, $W_{ud}$, and $W_{uq}$, respectively~\cite{heifets1997generalized}.

To establish a direct connection with the longitudinal impedance, we similarly expand it as a multivariate Taylor series,
\begin{equation}
    Z_\parallel(k;\mathbf{r}_{0\perp}, \mathbf{r}_{1\perp})
    =
    \sum_{ijkl}
    Z_{ijkl}^\parallel(k)
    x_0^i y_0^j x_1^k y_1^l.
    \label{eq:ZL-expansion}
\end{equation}
Substituting Eq.~\eqref{eq:ZL-expansion} into the Panofsky-Wenzel relation reveals that all transverse impedance components are fully determined by derivatives of the longitudinal impedance:
\begin{equation}
    Z_{xm}(k)=\frac{1}{k} Z_{0010}^\parallel(k), \qquad
    Z_{ym}(k)=\frac{1}{k} Z_{0001}^\parallel(k),
    \label{eq:ZT_mono}
\end{equation}
\begin{equation}
    Z_{xd}(k)=\frac{1}{k} Z_{1010}^\parallel(k), \qquad
    Z_{yd}(k)=\frac{1}{k} Z_{0101}^\parallel(k),
\label{eq:ZT_dip}
\end{equation}
\begin{equation}
    Z_{xq}(k)=\frac{2}{k} Z_{0020}^\parallel(k), \qquad
    Z_{yq}(k)=\frac{2}{k} Z_{0002}^\parallel(k).
\end{equation}
These relations are particularly useful in practical applications. They allow the transverse impedances to be derived directly from the longitudinal impedance and provide a stringent consistency check for impedance calculations. Moreover, the same relations apply to the transverse wake functions and transverse shunt impedances associated with different resonant modes.

The preceding formulation shows that, in a 3D asymmetric cavity, a single resonant mode can contribute simultaneously to several beam-coupling impedance components. For a given mode, the resonant wavenumber $k_r$, quality factor $Q$, and field pattern $(\mathbf E_0(\mathbf r),\mathbf B_0(\mathbf r))$ are intrinsic properties of the cavity mode. Therefore, all impedance components generated by this mode, including the coefficients obtained by expanding them with respect to the transverse offsets of the source and test particles, share the same $k_r$ and $Q$. What distinguishes the different longitudinal and transverse components is their coupling strength to the beam, which is described by the corresponding shunt-impedance-like factor. In other words, the same cavity mode may appear in different impedance components with a common resonance frequency and damping rate, but with different effective shunt impedances determined by the spatial structure of the mode fields and by the trajectories of the source and test particles. The intrinsic mode parameters $k_r$ and $Q$ can be obtained by several approaches, including wakefield simulations, eigenmode calculations, and RF measurements, whereas the shunt impedances must be evaluated for each specific impedance component.

\subsection{\label{sec:connectionToComputations} Connection to numerical wakefield and eigenmode computations}

As shown in the previous subsection, the wake function associated with an individual resonant mode can be formulated for $s>L_c$ and is directly related to the properties of that mode. On this basis, the total delta-function wake potential (following the terminology of~\cite{bane1984wake}) of a cavity can be decomposed into two parts:
\begin{equation}
    W_{u}^\mathrm{t}(s; \mathbf{r}_{0\perp}, \mathbf{r}_{1\perp})
    =
    W_{u}^\mathrm{S}(s; \mathbf{r}_{0\perp}, \mathbf{r}_{1\perp})
    +
    W_{u}^\mathrm{L}(s; \mathbf{r}_{0\perp}, \mathbf{r}_{1\perp})
\end{equation}
with
\begin{equation}
    W_{u}^\mathrm{L}(s; \mathbf{r}_{0\perp}, \mathbf{r}_{1\perp})
    =
    \sum_\mu
    W_{u}^\mu(s; \mathbf{r}_{0\perp}, \mathbf{r}_{1\perp})
\end{equation}
where $u=\parallel, x, y$. Here, $W_{u}^\mu(s; \mathbf{r}_{0\perp}, \mathbf{r}_{1\perp})$ denotes the contribution from the $\mu$-th resonant mode, as formulated in Eqs.~\eqref{eq:WakeFunctionLongitudinal} and~\eqref{eq:tra_wake_function_x}, which are valid for $s>L_c$. The term $W_{u}^\mathrm{S}(s; \mathbf{r}_{0\perp}, \mathbf{r}_{1\perp})$ represents the short-range wake, including contributions from non-resonant fields as well as the transient part of the resonant fields for $s<L_c$~\cite{weiland1980wake, bane1984wake}. Quantitatively, the short-range contribution satisfies
$W_{u}^\mathrm{S}(s; \mathbf{r}_{0\perp}, \mathbf{r}_{1\perp}) \rightarrow 0$ as $s$ becomes sufficiently large. This separation allows us to focus on the long-range resonant wakefields, which are of primary interest for the study of coupled-bunch instabilities. An analogous decomposition can also be introduced for the Taylor-expansion coefficients of the wake functions, although the corresponding formulation is omitted here for brevity.

In studies of coupled-bunch instabilities, the longitudinal monopolar wake and the transverse dipolar wakes are usually the most relevant components. They are defined as
\begin{equation}
    W_{\parallel 0}^\mathrm{t}(s)
    =
    W_\parallel^\mathrm{t}(s; \mathbf{0}, \mathbf{0}),
    \label{eq:WzMonopolarTotal}
\end{equation}
and
\begin{equation}
    W_{ud}^\mathrm{t}(s)
    =
    \left.
    \frac{\partial W_u^\mathrm{t}(s; \mathbf{r}_{0\perp}, \mathbf{r}_{1\perp})}
    {\partial u_0}
    \right|_{\substack{\mathbf{r}_{0\perp}=\mathbf{0}\\
                       \mathbf{r}_{1\perp}=\mathbf{0}}},
    \qquad u=x,y .
    \label{eq:WxyDipolarTotal}
\end{equation}
In numerical calculations of long-range wakefields, wake potentials are typically computed using a Gaussian driving bunch with codes such as CST Particle Studio~\cite{CST}, GdfidL, or ECHO3D. The resulting wake
potentials, denoted by $\mathcal{W}_{\parallel 0}^\mathrm{t}(s)$ and $\mathcal{W}_{ud}^\mathrm{t}(s)$, are the convolutions of the corresponding delta-function wake potentials with a Gaussian charge distribution of rms bunch length $\sigma_z$.

Considering the resonant-mode contributions to $\mathcal{W}_{\parallel 0}^\mathrm{t}(s)$ and $\mathcal{W}_{ud}^\mathrm{t}(s)$, analytical expressions are available, for example, in Sec.~3.2.4
of~\cite{chao2023handbook}. In the high-$Q$ limit and for
$s\gg \lambda_r \gg \sigma_z$, where $\lambda_r=2\pi/k_r$, the contribution from the $\mu$-th resonant mode can be approximated as
\begin{equation}
    \mathcal{W}_{\parallel 0}^\mu(s) \approx
    \frac{k_rcR_{\parallel 0}^{\mu}}{Q} e^{-\frac{1}{2}k_r^2\sigma_z^2}
    F_c'(s- \alpha \sigma_z^2),
    \label{eq:WzBunchApprox1}
\end{equation}
and
\begin{equation}
    \mathcal{W}_{ud}^\mu(s) \approx
    \frac{k_rcR_{ud}^\mu}{Q} e^{-\frac{1}{2}k_r^2\sigma_z^2}
    F_c(s- \alpha \sigma_z^2)
    \label{eq:WxyBunchApprox1}
\end{equation}
with $u=x,y$. Here, the parameters $k_r$, $Q$, and $\alpha$ correspond to the $\mu$-th mode, but the mode index is omitted for brevity. Comparison with Eqs.~\eqref{eq:WakeFunctionLongitudinal} and~\eqref{eq:tra_wake_function_x} shows that, in the long-range region, the bunch wake potentials reproduce the oscillatory behavior of the corresponding delta-function wake functions.

Equations~\eqref{eq:WzBunchApprox1} and~\eqref{eq:WxyBunchApprox1} can be used to fit numerically computed bunch wake potentials and thereby extract the parameters of a selected set of resonant modes, say $N$ modes. The extracted mode parameters can then be used for the analysis of coupled-bunch instabilities.

The wake-fitting approach was investigated recently in Ref.~\cite{yao2026particle}. An alternative approach is to first apply a fast Fourier transform (FFT) to the numerical wake data and then fit the resulting spectrum using the corresponding impedance model. This approach is motivated by the fact that the real parts of the impedance spectra of individual high-$Q$ modes are usually well separated from one another. Therefore, the frequency-domain representation enables rapid discrimination of distinct resonant modes, direct estimation of their resonant frequency ranges, and effective preliminary mode screening. Such operations improve computational efficiency and allow fast and robust fitting of the targeted mode parameters. In the following, we present the formalism required for such a fitting procedure.

Because the wakefields of high-$Q$ resonant modes decay slowly, numerical calculations of bunch wake potentials are usually truncated at a finite distance. As a result, the complete wake potentials or wake functions are not directly available from the simulation. However, we can construct impedance models corresponding to the truncated wake potentials or wake functions so that the fitting can still be performed using truncated wake potentials. For example, the monopolar wake function $W_{\parallel 0}^\mu(s)$ of the $\mu$-th resonant mode corresponding to Eq.~\eqref{eq:WakeFunctionLongitudinal} can be multiplied by a truncation function $H(s_1-s)$ with $H(x)$ the Heaviside function. This yields an equivalent impedance of
\begin{equation}
    Z_{\parallel 0}^{L 1} (k,s_1) =\sum_\mu Z_{\parallel 0}^{\mu 1} (k,s_1),
\label{eq:Zz0truncation}
\end{equation}
with
\begin{equation}
    Z_{\parallel 0}^{\mu 1} (k,s_1) = Z_{\parallel 0}^{\mu} (k) \tilde{T}_{\parallel}(k,s_1),
    \label{eq:Zz0truncation_single}
\end{equation}
where
\begin{equation}
    Z_{\parallel 0}^{\mu} (k) = \frac{R_{\parallel 0}^{\mu}}{1+iQ\left( \frac{k_r}{k}-\frac{k}{k_r}\right)}
    \label{eq:Zz0_single}
\end{equation}
and
\begin{align}
    \tilde{T}_{\parallel}(k,s_1)= & 1 - e^{-(\alpha-ik)s_1} \nonumber \\
    & \times \left[
    \cos(k_ds_1) - \frac{i(k_r^2-ik\alpha)}{k_dk} \sin (k_ds_1)
    \right].
    \label{eq:T_longth}
\end{align}
For high-$Q$ modes, $\alpha \ll k_d$ and $\tilde{T}_{\parallel}(k,s_1)$ can be approximated by
\begin{equation}
    \tilde{T}_{\parallel}(k,s_1) \approx
    1 - e^{-(\alpha-i k_\delta) s_1}
    \label{eq:TLsimple}
\end{equation}
where $k_\delta=k-k_d$.

Since the bunch wake potentials contains the short-range wakefields which are not suitable for fitting with the resonator impedance model, the wake potential over the interval $(s_1, s_2)$, with $s_2>s_1>L_c$, can be used to extract the mode parameters. In this case, the equivalent impedance is
\begin{align}
    Z_{\parallel 0}^{L 2} (k,s_1,s_2) =\sum_\mu Z_{\parallel 0}^{\mu 2} (k,s_1,s_2),
      \label{eq:Zz0FitModel}
\end{align}
with
\begin{align}
    Z_{\parallel 0}^{\mu 2} (k,s_1,s_2) = 
      & Z_{\parallel 0}^{\mu 1} (k,s_2) - Z_{\parallel 0}^{\mu 1} (k,s_1) \nonumber \\
      = & Z_{\parallel 0}^{\mu} (k) 
      \left[ \tilde{T}_{\parallel}(k,s_2) - \tilde{T}_{\parallel}(k,s_1) \right].
\end{align}
Since the long-range bunch wake potential is very close to the delta-function wake potential, Eq.~\eqref{eq:Zz0FitModel} can be directly used in fitting the computed impedance, as will be demonstrated in the following two sections.

The truncation formalism developed above extends directly to the transverse wake components. For the dipolar wake function $W_{ud}^{\mu}(s)$ ($u=x,y$) associated with the $\mu$-th resonant mode, truncation at $s=s_1$ yields the equivalent impedance
\begin{equation}
    Z_{u d}^{L 1} (k,s_1) =\sum_\mu Z_{u d}^{\mu 1} (k,s_1),
    \label{eq:Ztdtruncation}
\end{equation}
\begin{equation}
    Z_{u d}^{\mu 1} (k,s_1) = Z_{u d}^{\mu} (k) \tilde{T}_{u}(k,s_1),
\end{equation}
with $u=x,y$, and
\begin{equation}
    Z_{u d}^{\mu} (k) =\frac1 k \frac{R_{u d}^{\mu}}{1+iQ\left( \frac{k_r}{k}-\frac{k}{k_r}\right)},
    \label{eq:Ztd}
\end{equation}
\begin{equation}
    \tilde{T}_{u}(k,s_1)=1 - e^{-(\alpha-ik)s_1}
    \left[
    \cos(k_ds_1) - \frac{ik-\alpha}{k_d} \sin (k_ds_1)
    \right].
\end{equation}
Although the exact form of $\tilde T_u$ differs slightly from the longitudinal correction factor $\tilde T_\parallel$, both reduce in the high-$Q$ limit to
\begin{equation}
    \tilde{T}_{u}(k,s_1) \approx
    1 - e^{-(\alpha-i k_\delta) s_1}.
\end{equation}
Thus, longitudinal and transverse resonant impedances exhibit the same asymptotic correction due to wakefield truncation.

As in the longitudinal case, practical fitting should exclude the short-range wakefield region. Restricting the analysis to the interval $(s_1,s_2)$ therefore leads to
\begin{equation}
    Z_{u d}^{L 2} (k,s_1,s_2) =\sum_\mu Z_{u d}^{\mu 2} (k,s_1,s_2),
      \label{eq:ZtdFitModel}
\end{equation}
with
\begin{equation}
    Z_{u d}^{\mu 2} (k,s_1,s_2) = Z_{u d}^{\mu} (k) 
      \left[ \tilde{T}_{u}(k,s_2) - \tilde{T}_{u}(k,s_1) \right].
\end{equation}
Equation~\eqref{eq:ZtdFitModel} provides the fitting model used throughout the remainder of this work for extracting transverse resonant-mode parameters from truncated wakefield data.

In contrast to the wakefield calculations, the parameters of specific resonant modes of RF cavities can be computed using eigenmode solvers such as CST and GdfidL. For a given cavity geometry, eigenmode solvers typically determine the resonant modes by solving the discretized source-free Maxwell's equations with the specified boundary conditions. The resonant frequency is obtained from the eigenvalue of each mode. The quality factor is evaluated from the ratio of the stored electromagnetic energy to the power loss per RF period, including wall and/or material losses when they are included in the model. The longitudinal shunt impedance is then calculated from the accelerating voltage seen by a relativistic test particle, normalized by the dissipated power; equivalently, the solver first obtains the normalized shunt impedance (i.e., $R_s/Q$) from the mode fields and stored energy, and the shunt impedance is obtained as $R_s=(R_s/Q)*Q$. In this way, the resonant frequency, quality factor, and shunt impedance are consistently extracted from the same numerically computed eigenmode fields.

An eigenmode solver evaluates the coupling of a given mode to a prescribed particle trajectory by integrating the modal fields along that trajectory. The resulting accelerating voltage has the same form as Eq.~\eqref{eq:AccVoltageSource}, with the transverse position of the trajectory taken into account. Once the modal field pattern, stored energy $U_0$, resonant frequency $\omega_r$, and quality factor $Q$ are known, the longitudinal shunt impedance can be obtained from the offset-dependent accelerating voltage as
\begin{equation}
    \mathcal{R}_\parallel(\mathbf{r}_{0\perp})=
    \frac{Q}{\omega_r U_0}
    \left| \mathcal{V}_0 (\mathbf{r}_{0\perp}) \right|^2.
    \label{eq:RLeigenmode}
\end{equation}
Here the offset dependence is expressed through a single trajectory coordinate $\mathbf{r}_{0\perp}$, in contrast to the generalized shunt impedance $R_\parallel$ in Eq.~\eqref{eq:RLgeneral}, which depends on the transverse offsets of both the source and test particles.

For the longitudinal monopolar impedance, corresponding to $Z_{0000}^\parallel$ in Eq.~\eqref{eq:ZL-expansion}, the associated shunt impedance is obtained by evaluating Eqs.~\eqref{eq:RLgeneral} and~\eqref{eq:RLeigenmode} on axis:
\begin{equation}
    R_{\parallel 0} = R_\parallel(\mathbf{0},\mathbf{0})=\mathcal{R}_\parallel(\mathbf{0}).
    \label{eq:cal_Rs_eigenmode}
\end{equation}
The transverse monopolar component can be derived from $\mathcal{R}_\parallel$ as
\begin{equation}
    R_{um} = 
    \left. \frac{\partial R_\parallel}{\partial u_1} \right|_{\substack{\mathbf{r}_{0\perp}=\mathbf{0}\\
            \mathbf{r}_{1\perp}=\mathbf{0}}}
    =
    \left. \frac{1}{2} \frac{\partial \mathcal{R}_\parallel}{\partial u_0} \right|_{\mathbf{r}_{0\perp}=\mathbf{0}}.
    \label{eq:cal_Rs_eigenmode_m}
\end{equation}
The transverse dipolar component is obtained from the squared first derivative of the modal voltage,
\begin{equation}
    R_{ud} =
    \left. \frac{\partial^2 R_\parallel}{\partial u_1 \partial u_0} \right|_{\substack{\mathbf{r}_{0\perp}=\mathbf{0}\\
            \mathbf{r}_{1\perp}=\mathbf{0}}}
    =\frac{Q}{\omega_r U_0}
    \left. \left| \frac{\partial \mathcal{V}_0}{\partial u_0} \right|^2 \right|_{\mathbf{r}_{0\perp}=\mathbf{0}}.
    \label{eq:cal_Rs_eigenmode_d}
\end{equation}
Note that $R_{ud}$ cannot be directly computed from $\mathcal{R}_\parallel$. Finally, the transverse quadrupolar component is related to the second-order variation of $\mathcal{R}_\parallel$ after subtracting the dipolar contribution,
\begin{equation}
    R_{uq} = \left. \frac{1}{2} \frac{\partial^2 \mathcal{R}_\parallel}{\partial u_0^2} \right|_{\mathbf{r}_{0\perp}=\mathbf{0}}
    - R_{ud}.
    \label{eq:cal_Rs_eigenmode_q}
\end{equation}
Here $u=x$ or $y$ corresponds to the horizontal or vertical transverse impedance component, respectively. These expressions show explicitly how the longitudinal and transverse shunt impedances can be extracted from the spatial dependence of the eigenmode voltage.

Equations~\eqref{eq:cal_Rs_eigenmode_m}-\eqref{eq:cal_Rs_eigenmode_q} establish a direct connection between eigenmode calculations and the generalized impedance decomposition. They show that the transverse impedance strengths can be obtained from the transverse dependence of the longitudinal accelerating voltage, without performing separate transverse wakefield simulations. This result follows from the Panofsky-Wenzel relation: the transverse voltage is determined by the transverse gradient of the longitudinal voltage, and therefore the transverse impedance components are encoded in the spatial variation of $\mathcal{V}_0(\mathbf r_{0\perp})$.

The expressions of Eqs.~\eqref{eq:cal_Rs_eigenmode_m}-\eqref{eq:cal_Rs_eigenmode_q} are generally consistent with the formulation of Ref.~\cite{arsenyev2019method}. One exception concerns the dipolar shunt impedance, for which Ref.~\cite{arsenyev2019method} gives, in the notation of the present paper,
\begin{equation}
    R_{ud}^{AS} =
    \left. \frac{1}{4\mathcal{R}_\parallel}
    \left( \frac{\partial \mathcal{R}_\parallel}{\partial u_0} \right)^2
    \right|_{\mathbf{r}_{0\perp}=\mathbf{0}}
    = \frac{Q}{\omega_r U_0}
    \left. \left( \frac{\partial \left|\mathcal{V}_0\right|}{\partial u_0} \right)^2 \right|_{\mathbf{r}_{0\perp}=\mathbf{0}},
    \label{eq:cal_Rs_eigenmode_d_arsenyev}
\end{equation}
which is consistent with the driving impedance $Z_x^\mathrm{driv}$ in Eq.~(13) of Ref.~\cite{arsenyev2019method}. Equation~\eqref{eq:cal_Rs_eigenmode_d_arsenyev} should be regarded as an approximation to Eq.~\eqref{eq:cal_Rs_eigenmode_d}, rather than as a generally equivalent form. The distinction is that Eq.~\eqref{eq:cal_Rs_eigenmode_d} involves the modulus squared of the complex voltage derivative, $\left|\partial \mathcal{V}_0/\partial u_0\right|^2$, whereas Eq.~\eqref{eq:cal_Rs_eigenmode_d_arsenyev} uses the squared derivative of the voltage amplitude, $\left(\partial |\mathcal{V}_0|/\partial u_0\right)^2$. These two quantities are identical only when the transverse variation of $\mathcal{V}_0$ changes its amplitude but not its phase. A sufficient condition for this approximation to hold is that the longitudinal eigenfield in Eq.~\eqref{eq:AccVoltageSource} can be separated as
$E_{0z}(x_0,y_0,z)=A(x_0,y_0)B(z)$,
so that the transverse offset modifies only the overall amplitude of the accelerating voltage. This separability may be a good approximation for simple or highly symmetric cavity geometries. For general 3D asymmetric cavities, however, the longitudinal field can have a non-separable dependence on transverse position and longitudinal coordinate, leading to an offset-dependent phase of $\mathcal{V}_0$. In such cases, Eq.~\eqref{eq:cal_Rs_eigenmode_d} should be used for the dipolar shunt impedance.

\section{\label{sec:validation}Validation Using an Axisymmetric Pillbox Cavity}

Before applying the formulation to realistic RF cavities, we first benchmark it with a simple cavity geometry for which analytical results are available. Specifically, we consider an axisymmetric pillbox cavity and compare the resonant impedances obtained from the analytical theory with those extracted from wakefield and eigenmode simulations. This comparison tests both the resonant-impedance expressions and the truncation-correction scheme developed in Sec.~\ref{sec:theory}. The agreement among the three approaches provides a stringent validation of the framework and establishes a reference case for the more realistic cavity studies presented in the following section.

\subsection{\label{sec:ValidateLongitudinal}Validation of longitudinal resonant impedances}

For an axisymmetric pillbox cavity with perfectly conducting walls, analytical expressions for the resonant impedances can be obtained directly from the cavity eigenfields. The electromagnetic fields of a TM$_{mnp}$ mode in cylindrical coordinates are given by
\begin{equation}
\begin{aligned}
B_{0r} &= \frac{2iE_0m{k_{r}}R^2}{cr v_{mn}^2} 
         \cos(m\theta+\varphi_0)\cos\left(\frac{p\pi}{L_c}z\right)J_m\left(\frac{v_{mn}}{R}r\right), \\[1ex]
B_{0\theta} &= \frac{-2iE_0{k_{r}}R^2}{c v_{mn}^2}
              \sin(m\theta+\varphi_0)\cos\left(\frac{p\pi}{L_c}z\right)
              \mathcal{R} (r), \\[2ex]
E_{0r} &= \frac{-2E_0p\pi R^2}{L_c v_{mn}^2}
          \sin(m\theta+\varphi_0) \sin\left(\frac{p\pi}{L_c}z\right)
          \mathcal{R} (r), \\[1ex]
E_{0\theta} &= \frac{-2E_0mp\pi R^2}{L_c r v_{mn}^2}
              \cos(m\theta+\varphi_0)             \sin\left(\frac{p\pi}{L_c}z\right)
              J_m\left(\frac{v_{mn}}{R}r\right), \\[1ex]
E_{0z} &= 2E_0\sin(m\theta+\varphi_0)\cos\left(\frac{p\pi}{L_c}z\right)
          J_m\left(\frac{v_{mn}}{R}r\right),
\end{aligned}
\label{eq:PB_cav_field}
\end{equation}
where 
\begin{equation}
    \mathcal{R} (r) =\left[\frac{m}{r}J_m\left(\frac{v_{mn}}{R}r\right)
              -\frac{v_{mn}}{R}J_{m+1}\left(\frac{v_{mn}}{R}r\right)\right]
\end{equation}
with $k_{r}^2 = \left( \frac{v_{mn}}{R} \right)^2 + \left( \frac{\pi p}{g} \right)^2$. Here $m$, $n$, and $p$ denote the azimuthal, radial, and longitudinal mode indices, respectively. $\varphi_0$ is an arbitrary constant. $L_c$ and $R$ are the cavity length and radius, and $v_{mn}$ is the $n$-th root of the $m$-th order Bessel function ${J}_m(x)$.

Substituting the analytical eigenfields of Eq.~\eqref{eq:PB_cav_field} into the general expression for the longitudinal resonant impedance, Eq.~\eqref{eq:longitudinal impedance}, yields the corresponding normalized shunt impedances $R_{\parallel}/Q$~\cite{weiland1980wake, Bane2012Expressions}. For azimuthally symmetric modes (i.e., $m=0$),
\begin{equation}
\begin{aligned}
\frac{R_{\parallel}(\mathbf{r}_{0\perp}, \mathbf{r}_{1\perp})}{Q}
=
\frac{Z_0  J_0\left(\frac{v_{0n}}{R} r_1\right) J_0\left(\frac{v_{0n}}{R} r_0\right) }{v_{0n}^2 \left(J_0'(v_{0n})\right)^2} \mathcal{K} (k_r,p);
\end{aligned}
\label{eq:PB_cav_m0Rs/Q}
\end{equation}
whereas for modes with $m\neq0$,
\begin{equation}
\begin{aligned}
\frac{R_{\parallel}(\mathbf{r}_{0\perp}, \mathbf{r}_{1\perp})}{Q}
=& 2\sin(m\theta_1+\varphi_0)\sin(m\theta_0+\varphi_0) \\
&\frac{Z_0 J_m\left(\frac{v_{mn}}{R} r_1\right) J_m\left(\frac{v_{mn}}{R} r_0\right)}{v_{mn}^2\left(J_m'(v_{mn})\right)^2} 
\mathcal{K} (k_r,p)
\label{eq:PB_cav_m123Rs/Q}
\end{aligned}
\end{equation}
where 
\begin{equation}
    \mathcal{K} (k,p) =\frac{8}{\pi L_c k}
\begin{cases}
\dfrac{\sin^2\left(\dfrac{kL_c}{2}\right)}{1+\delta_{0p}} & \text{for even } p \\[1.5em]
\cos^2\left(\dfrac{kL_c}{2}\right) & \text{for odd } p
\end{cases}.
\end{equation}
A direct consequence of Eqs.~\eqref{eq:PB_cav_m0Rs/Q} and~\eqref{eq:PB_cav_m123Rs/Q} is that, for $m\neq 0$, the on-axis longitudinal normalized shunt impedance of the axisymmetric pillbox cavity vanishes, i.e.,
$R_{\parallel 0}^{\mu}/Q = R_{\parallel}(\mathbf 0,\mathbf 0)/Q=0$.
Thus, in an ideal axisymmetric pillbox cavity, higher-order azimuthal modes do not couple longitudinally to an on-axis beam.

\begin{figure}[!h]
\includegraphics{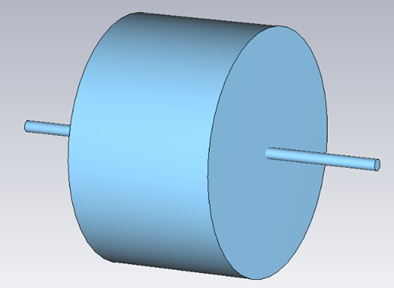}
\caption{\label{fig:fig_01_pill_box} CST model of the pillbox-like cavity. The main dimensions are listed in Table~\ref{tab:table1_par_PB}.}
\end{figure}

To benchmark the analytical predictions against numerical calculations, a corresponding pillbox cavity model was constructed in CST, as shown in Fig.~\ref{fig:fig_01_pill_box}. The analytical model assumes an ideal closed pillbox cavity with perfectly conducting walls, whereas the CST model includes beam pipes with radii much smaller than the cavity radius in order to enable wakefield simulations. The main cavity dimensions are summarized in Table~\ref{tab:table1_par_PB}.
\begin{table}[!t]
\caption{\label{tab:table1_par_PB}Geometrical parameters of the pillbox cavity.}
\begin{ruledtabular}
\begin{tabular}{cccc}
\multicolumn{2}{c}{\textrm{cavity}}&
\multicolumn{2}{c}{\textrm{beam pipe}}\\
length $g$ &radius $R$ &length $L$ &radius $r$ \\%
$[\text{mm}]$ & $[\text{mm}]$ & $[\text{mm}]$ & $[\text{mm}]$ \\ \hline%
260        &210     &200          &10 \\
\end{tabular}
\end{ruledtabular}
\end{table}

The presence of beam pipes introduces a small but finite deviation from the ideal closed-cavity boundary condition. As a result, the simulated eigenmodes are not exactly identical to the analytical pillbox modes. This difference is expected to be weak for well-confined low-frequency modes, but becomes more visible for higher-frequency modes whose fields extend more strongly into the beam pipes. Consequently, discrepancies between the simulated and analytical values of $R_{\parallel0}^{\mu}/Q$ tend to increase at higher frequencies.

We apply the proposed fitting procedure to the longitudinal impedance obtained from truncated wake potentials of the pillbox cavity. According to Eq.~\eqref{eq:Zz0FitModel}, the truncation length is chosen to exceed both the cavity length and the bunch length such that the residual wakefield is dominated by the resonant contribution rather than short-range transients.
The choice of truncation lengths is guided by the modal damping coefficient $\alpha = k_r/(2Q)$ in Eqs.~\eqref{eq:Zz0FitModel} and~\eqref{eq:ZtdFitModel}. For GHz-range resonances with $Q \gtrsim 10^3$, typical of storage-ring RF cavities, the corresponding damping length exceeds $100~\mathrm{m}$.
To obtain well-resolved resonant signatures from the difference between two truncated impedance spectra, a sufficiently large separation between the truncation lengths is required to ensure stable cancellation of the common short-range contribution and clear resolution of the resonant peaks. In this study, we consider $L_1 = 300~\mathrm{m}$ and $L_2 = 900~\mathrm{m}$ as representative choices.

\begin{figure}
\includegraphics[width=8cm]{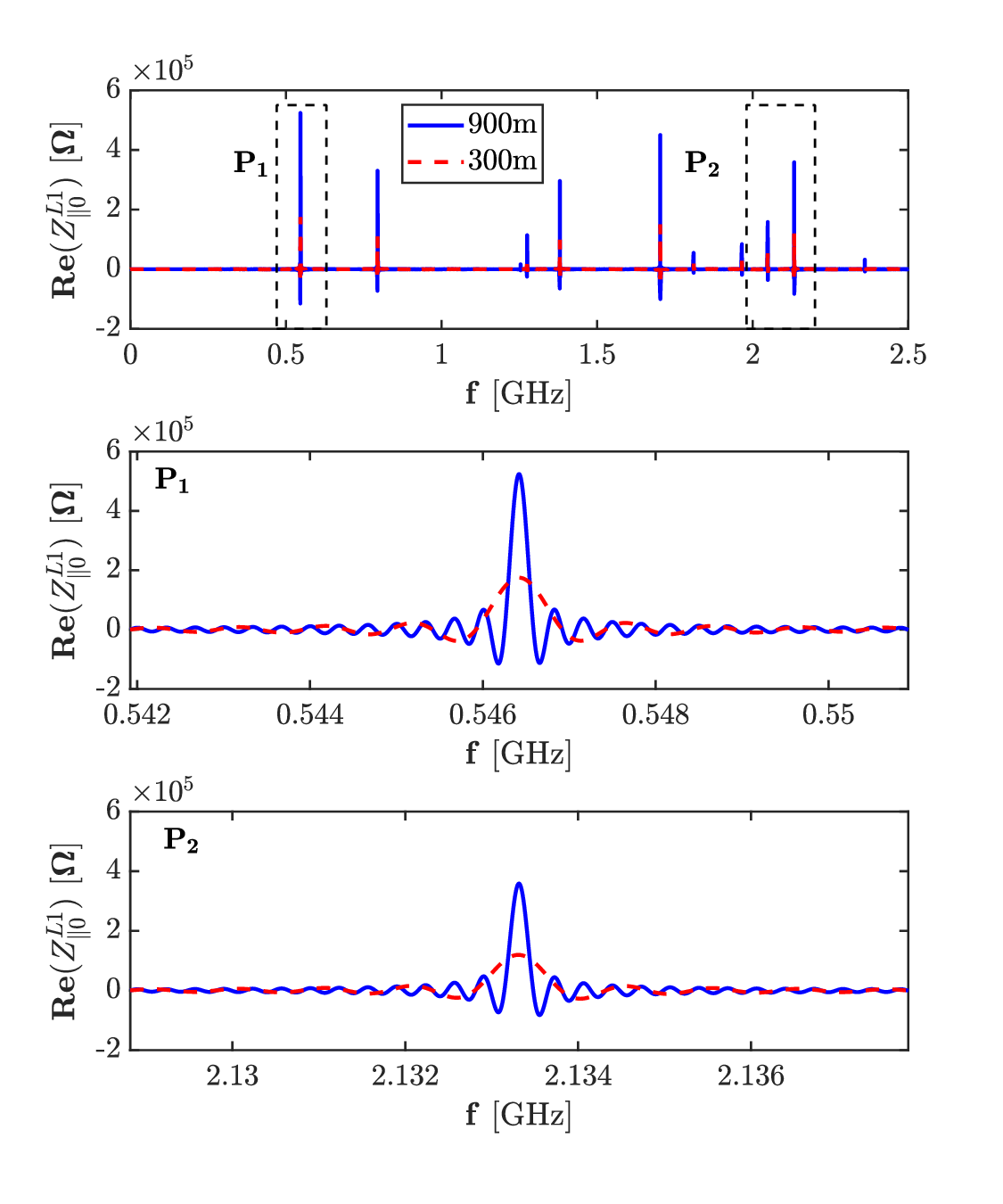}
\caption{\label{fig:fig_03_L_truncted_impedance_PB}Real part of the longitudinal impedance obtained from wake potentials truncated at $L_1=300$ m and $L_2=900$ m in the pillbox cavity. Top: full spectrum up to $2.5~\mathrm{GHz}$. Middle: zoomed-in view around $0.546~\mathrm{GHz}$, corresponding to the fundamental mode. Bottom: zoomed-in view around $2.133~\mathrm{GHz}$.}
\end{figure}

Figure~\ref{fig:fig_03_L_truncted_impedance_PB} shows the corresponding real parts of the longitudinal impedance up to $2.5~\mathrm{GHz}$, as calculated with CST. As predicted by Eq.~\eqref{eq:Zz0truncation}, wakefield truncation produces oscillatory distortions around the resonance peaks. These distortions originate from the abrupt termination of wakefields that have not fully decayed within the simulated wake length. As the truncation length increases, the oscillations become more rapidly modulated, while their envelope decreases. In the limit of a sufficiently long length, the truncated impedance converges to the ideal resonant impedance.

\begin{figure}
\includegraphics[width=8cm]{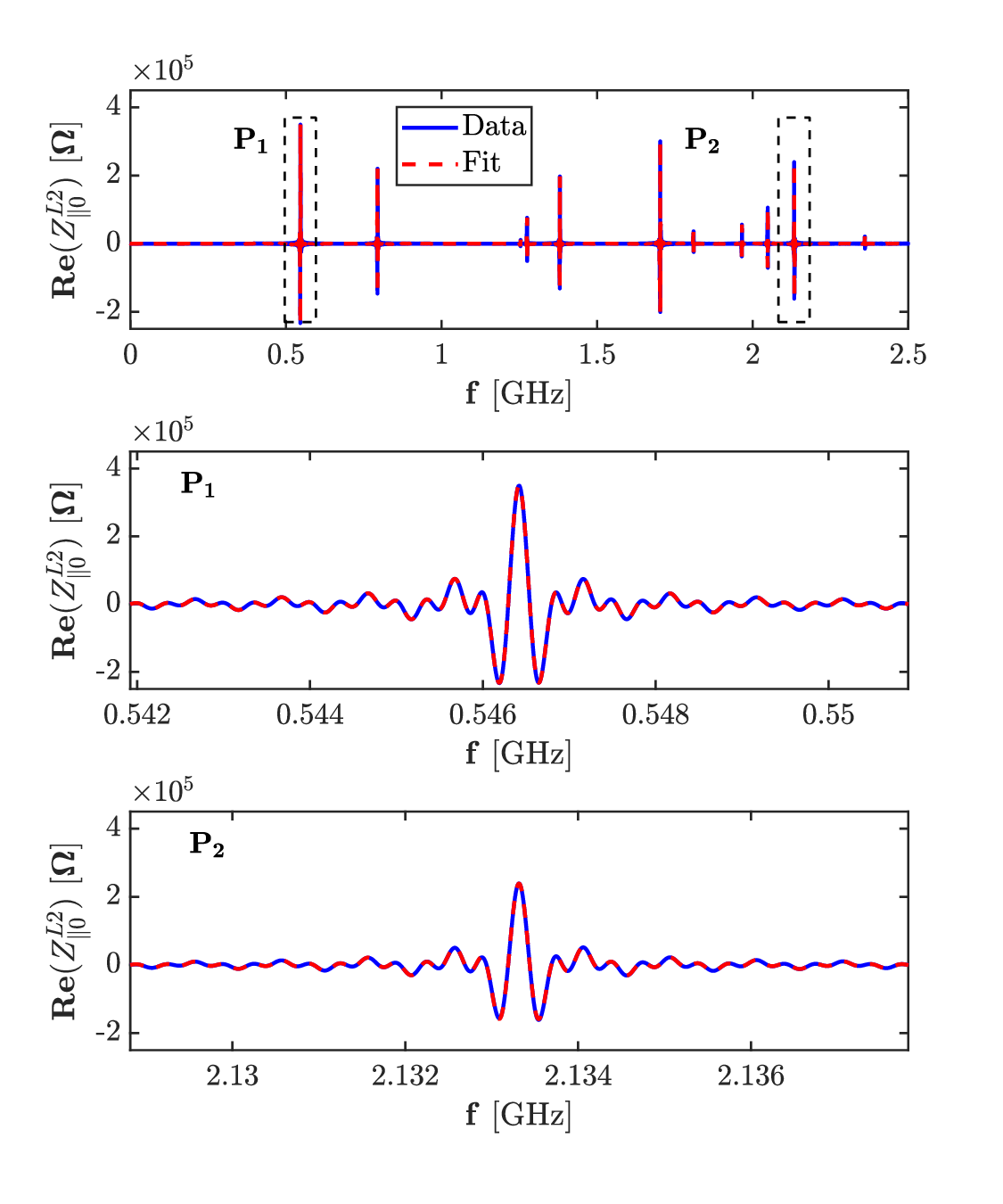}
\caption{\label{fig:fig_04_L_impedance_fit_PB}Difference between the longitudinal impedance spectra obtained with truncation lengths of $300~\mathrm{m}$ and $900~\mathrm{m}$ (blue dots), together with the fit based on Eq.~\eqref{eq:Zz0FitModel} (red solid line). Top: full spectrum up to $2.5~\mathrm{GHz}$. Middle: zoomed-in view around $0.546~\mathrm{GHz}$, corresponding to the fundamental mode. Bottom: zoomed-in view around $2.133~\mathrm{GHz}$.}
\end{figure}

To suppress the common short-range wakefield contribution and isolate the truncation-dependent resonant component described by Eq.~\eqref{eq:Zz0FitModel}, the difference between the impedance spectra obtained with truncation lengths of $300~\mathrm{m}$ and $900~\mathrm{m}$ was used as the input to the fitting procedure. The resulting fit is compared with the impedance difference in Fig.~\ref{fig:fig_04_L_impedance_fit_PB}. Excellent agreement is obtained after iterative optimization, demonstrating that the proposed fitting model accurately reproduces the truncation-dependent impedance variation predicted by the theory.

Table~\ref{tab:table2_fit_data_Lcompare_PB} compares the resonant frequencies $f_r$ and normalized shunt impedances $R_{\parallel0}^{\mu}/Q$ extracted from the proposed fitting procedure with the analytical values obtained from Eq.~\eqref{eq:PB_cav_m0Rs/Q}. Good agreement is obtained for the low- and intermediate-frequency modes. The discrepancies gradually increase at higher frequencies, where the finite beam pipes included in the CST model perturb the ideal pillbox eigenmodes assumed in the analytical solution.
\begin{table*}
\caption{\label{tab:table2_fit_data_Lcompare_PB}Comparison of resonant frequencies $f_r$ and normalized shunt impedances $R_{\parallel0}^{\mu}/Q$ obtained from the fitting procedure and analytical calculations for the pillbox cavity.}
\begin{ruledtabular}
\begin{tabular}{ccccc}%
\multicolumn{2}{c}{Truncated impedance fitting} & \multicolumn{3}{c}{Theoretical calculations} \\%
Frequency & $R_{\parallel 0}^{\mu}/Q$ & $m\ n\ p$ & Frequency & $R_{\parallel 0}^{\mu}/Q$ \\%
$[\text{GHz}]$ & $[\Omega]$ &  & $[\text{GHz}]$ & $[\Omega]$ \\%
\hline%
0.546 & 103 & 010 & 0.546 & 103 \\
0.794 & 44.5  & 011 & 0.794 & 44.5  \\
1.254& 1.42   & 020 & 1.254& 1.47   \\
1.275& 9.54   & 012 & 1.276& 9.57   \\
1.380& 22.9  & 021 & 1.380& 24.0  \\
1.703& 28.2  & 022 & 1.703& 29.2  \\
1.810& 3.20   & 013 & 1.814& 3.23    \\
1.965& 4.58   & 030 & 1.966& 5.19   \\
2.048& 8.27   & 031 & 2.049& 9.11   \\
2.133& 18.0  & 023 & 2.137& 18.7  \\
2.361& 1.47   & 014 & 2.370& 1.43   \\
\end{tabular}
\end{ruledtabular}
\end{table*}
\begin{figure}
\includegraphics[width=8cm]{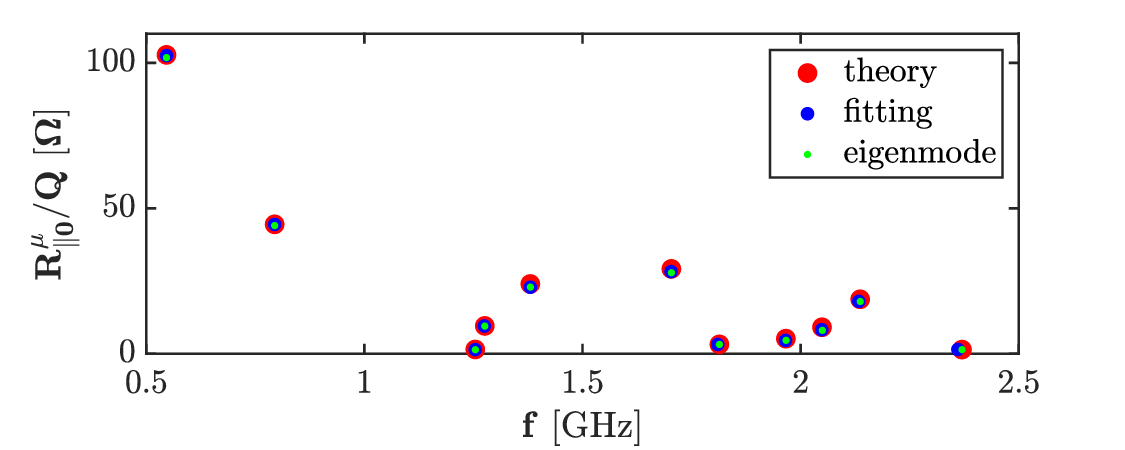}
\caption{\label{fig:fig_05_L_RsQ_compare_PB}Comparison of resonant frequencies $f_r$ and normalized shunt impedances $R_{\parallel0}^{\mu}/Q$ obtained from analytical calculations, the fitting procedure, and CST eigenmode simulations for the pillbox cavity.}
\end{figure}
Figure~\ref{fig:fig_05_L_RsQ_compare_PB} further compares the modal parameters obtained from the analytical model, the fitting procedure, and direct CST eigenmode calculations. The fitted resonant frequencies and $R_{\parallel0}^{\mu}/Q$ values closely reproduce the eigenmode results over the entire frequency range and remain in good agreement with the analytical predictions. The systematic deviation observed at higher frequencies appears in both the fitting and eigenmode results, indicating that it originates from the finite beam-pipe geometry included in the CST model rather than from the fitting procedure itself.

\subsection{\label{sec:ValidateTransverse}Validation of transverse resonant impedances}

Applying Eq.~\eqref{eq:ZT_mono} to Eqs.~\eqref{eq:PB_cav_m0Rs/Q} and \eqref{eq:PB_cav_m123Rs/Q} in the pillbox cavity solution shows that the transverse monopolar impedance vanishes identically on axis. This result is consistent with the symmetry argument developed in Sec.~\ref{sec:theory} and confirms that no transverse monopolar impedance can arise in an axisymmetric cavity.

The transverse dipolar impedance, which is of primary importance for beam stability analysis, can be obtained from Eq.~\eqref{eq:ZT_dip} by evaluating the mixed transverse derivatives of Eqs.~\eqref{eq:PB_cav_m0Rs/Q} and \eqref{eq:PB_cav_m123Rs/Q}. For azimuthally symmetric modes ($m=0$), the resulting dipolar shunt impedance vanishes on the cavity axis ($r=0$), consistent with Eq.~\eqref{eq:cal_Rs_eigenmode_d}, which predicts that $m=0$ modes do not contribute to the dipolar transverse impedance of an axisymmetric cavity. More generally, the dipolar impedance also vanishes on axis for all modes with azimuthal order $m\ge2$. Consequently, only the $m=1$ modes generate a finite dipolar transverse impedance for an on-axis beam. Taking the limit $r\rightarrow0$ and $\theta\rightarrow0$ yields
\begin{equation}
\begin{aligned}
\frac{ R_{u d}^{\mu}}{Q }
= \frac{Z_0}{2 R^2 \left(J_m'(v_{mn})\right)^2} \mathcal{K} (k_r,p).
\end{aligned}
\end{equation}

\begin{figure}
\includegraphics[width=8cm]{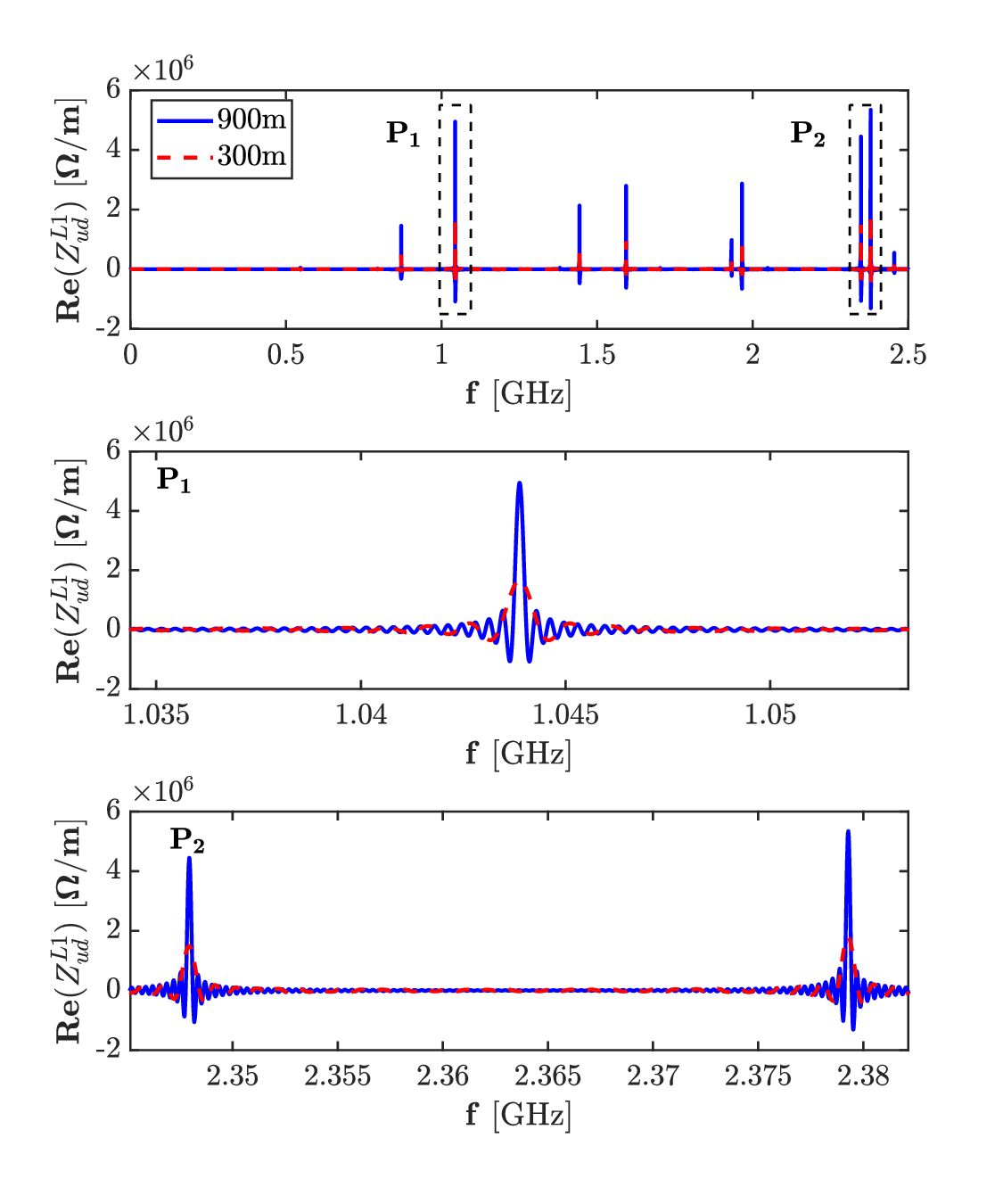}
\caption{\label{fig:fig_06_T_truncted_impedance_PB}Real part of the transverse dipolar impedance obtained from wake potentials truncated at $L_1=300~\mathrm{m}$ and $L_2=900~\mathrm{m}$ in the pillbox cavity. Top: full spectrum up to $2.5~\mathrm{GHz}$. Middle: zoomed-in view around $1.044~\mathrm{GHz}$, corresponding to the fundamental mode. Bottom: zoomed-in view around $2.365~\mathrm{GHz}$.}
\end{figure}

To validate the transverse fitting model of Eq.~\eqref{eq:ZtdFitModel}, the same wakefield truncation lengths, $L_1=300~\mathrm{m}$ and $L_2=900~\mathrm{m}$, are adopted. The transverse dipolar impedance is extracted from wakefield simulations using Eq.~\eqref{eq:ZT_dip}, which requires evaluating the response for two opposite transverse beam offsets and normalizing the resulting impedance difference by the applied displacement. To reduce the influence of numerical noise while remaining within the linear regime, transverse bunch offsets of $\pm1~\mathrm{mm}$ are used throughout the calculation. This value is sufficiently small compared with the cavity dimensions listed in Table~\ref{tab:table1_par_PB}, while providing adequate numerical sensitivity for the extraction of the dipolar impedance. The resulting offset-normalized transverse dipolar impedance spectra obtained from CST for the two truncation lengths are shown in Fig.~\ref{fig:fig_06_T_truncted_impedance_PB}. Similar to the longitudinal case, finite wakefield truncation introduces oscillatory distortions around the resonance peaks, whose amplitudes decrease as the truncation length increases.

\begin{figure}
\includegraphics[width=8cm]{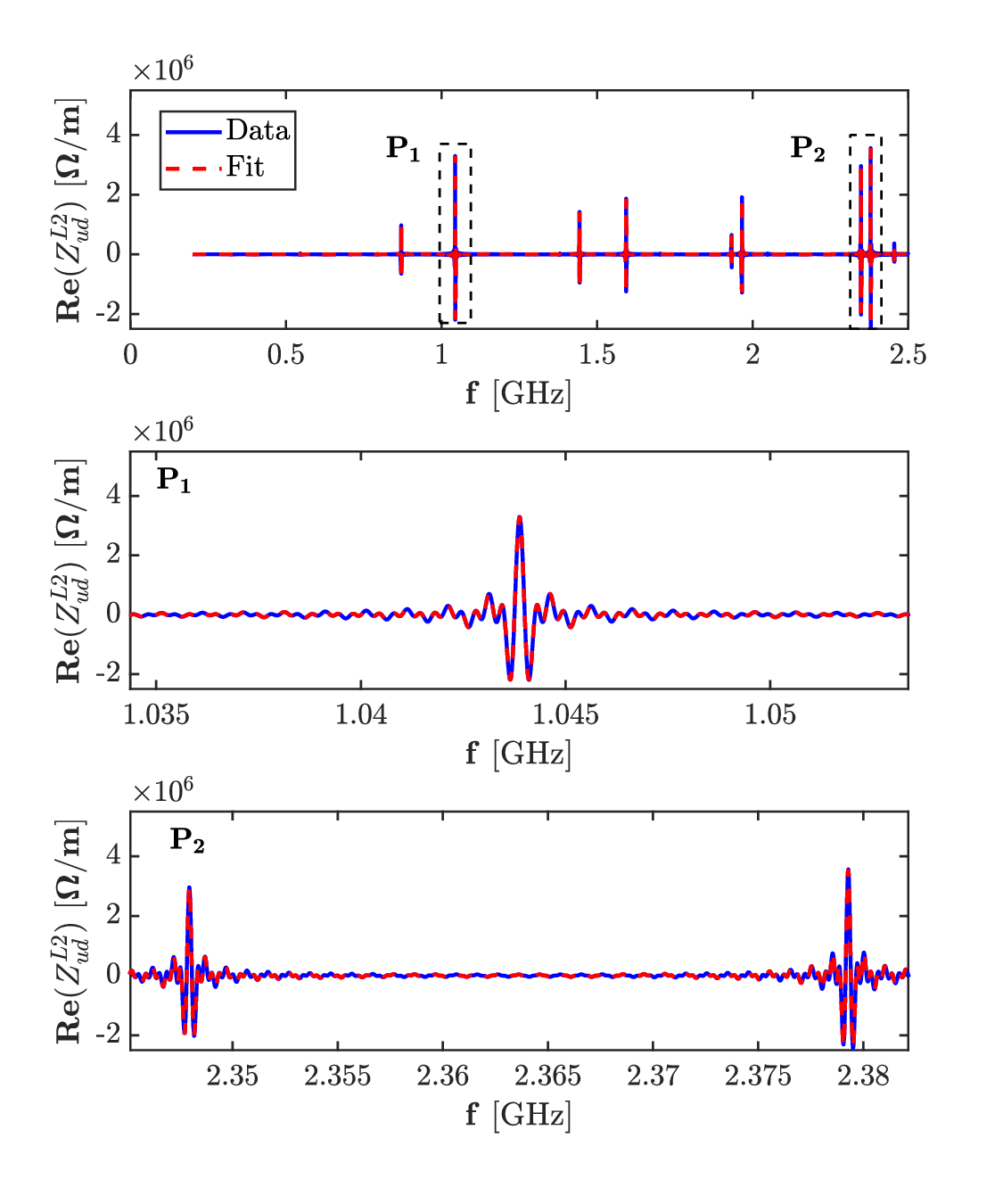}
\caption{\label{fig:fig_07_T_impedance_fit_PB}Difference between the transverse impedance spectra obtained with truncation lengths of $300~\mathrm{m}$ and $900~\mathrm{m}$ (blue dots), together with the fit based on Eq.~\eqref{eq:ZtdFitModel} (red solid line). Top: full spectrum up to $2.5~\mathrm{GHz}$. Middle: zoomed-in view around $1.044~\mathrm{GHz}$, corresponding to the fundamental mode. Bottom: zoomed-in view around $2.365~\mathrm{GHz}$.}
\end{figure}

The transverse impedance difference spectrum obtained from Fig.~\ref{fig:fig_06_T_truncted_impedance_PB} was subsequently fitted using Eq.~\eqref{eq:ZtdFitModel}. The resulting fit is compared with the CST data in Fig.~\ref{fig:fig_07_T_impedance_fit_PB}. Excellent agreement is observed over the entire frequency range after iterative optimization, indicating that the proposed fitting model accurately reproduces the truncation-dependent behavior of the transverse resonant impedance.

Table~\ref{tab:table3_fit_T_data_compare_PB} compares the resonant frequencies $f_r$ and normalized dipolar shunt impedances $R_{ud}^{\mu}/Q$ extracted from the proposed fitting procedure with the corresponding analytical predictions. Overall, excellent agreement is observed for the dominant dipolar modes. Similar to the longitudinal case, the discrepancies become more pronounced at higher frequencies, where the finite beam pipes included in the CST model perturb the ideal eigenmodes assumed in the analytical pillbox solution.
\begin{table*}
\caption{\label{tab:table3_fit_T_data_compare_PB}Comparison of resonant frequencies $f_r$ and normalized dipolar shunt impedances $R_{ud}^{\mu}/Q$ obtained from the fitting procedure and analytical calculations for the pillbox cavity.}
\begin{ruledtabular}
\begin{tabular}{ccccc}%
\multicolumn{2}{c}{Truncated impedance fitting} & \multicolumn{3}{c}{Theoretical calculations} \\
Frequency & $R_{u d}^{\mu}/{Q }$ & $m\ n\ p$ & Frequency & $R_{u d}^{\mu}/{Q }$ \\
$[\text{GHz}]$ & $[\Omega/\text{m}^2]$ &  & $[\text{GHz}]$ & $[\Omega/\text{m}^2]$ \\
\hline%
0.871  & 3280  & 120 & 0.871  & 3420  \\
1.044 & 11100 & 121 & 1.044 & 10800 \\
1.444 & 4800  & 122 & 1.445 & 4340  \\
1.593 & 6290  & 130 & 1.594 & 6050  \\
1.932 & 2190  & 123 & 1.936 & 1810  \\
1.966 & 6480  & 132 & 1.967 & 7170  \\
2.348 & 10100 & 133 & 2.352 & 9280  \\
2.379 & 12100 & 141 & 2.382 & 12900 \\
2.455 & 1270  & 124 & 2.465 & 879   \\
\end{tabular}
\end{ruledtabular}
\end{table*}

\begin{figure}
\includegraphics[width=8cm]{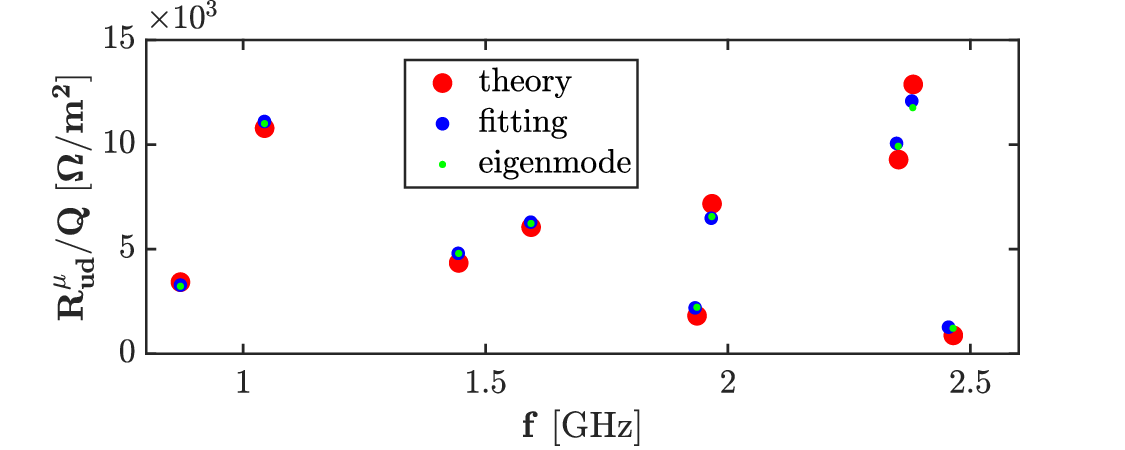}
\caption{\label{fig:fig_08_T_RsQ_compare_PB}Comparison of resonant frequencies $f_r$ and normalized dipolar shunt impedances $R_{ud}^{\mu}/Q$ obtained from analytical calculations, the fitting procedure, and CST eigenmode simulations for the pillbox cavity.}
\end{figure}

Figure~\ref{fig:fig_08_T_RsQ_compare_PB} further compares the modal parameters obtained from the analytical model, the fitting procedure, and direct CST eigenmode calculations. The fitted resonant frequencies and $R_{ud}^{\mu}/Q$ values closely reproduce the eigenmode results across the entire frequency range and remain in good agreement with the analytical predictions. The systematic deviation observed at higher frequencies appears in both the fitting and eigenmode results, indicating that it originates from the finite beam-pipe geometry included in the CST model rather than from the fitting procedure itself. The agreement among the three independent approaches provides a stringent validation of the proposed transverse resonant-impedance extraction method.

The foregoing comparisons demonstrate that the proposed fitting procedure accurately extracts the resonant frequencies $f_r$, longitudinal shunt impedances $R_{\parallel0}^{\mu}/Q$, and transverse dipolar shunt impedances $R_{ud}^{\mu}/Q$ from truncated wakefield simulations. Excellent agreement with both analytical predictions and CST eigenmode calculations is obtained for the axisymmetric pillbox cavity, thereby validating the proposed resonant-impedance extraction framework.

A noteworthy feature of Eqs.~\eqref{eq:Zz0FitModel} and~\eqref{eq:ZtdFitModel} is that, in the high-$Q$ limit relevant to weakly damped cavity modes, the truncation functions depend only weakly on the quality factor. In this limit, $\alpha\approx 0$ and $k_d\approx k_r$, giving $\tilde{T}_{\parallel}(k,s)\approx \tilde{T}_{u}(k,s)\approx 1-e^{-ik_r s}$. Consequently, the truncated impedance spectrum is governed mainly by the resonant frequency and normalized shunt impedance, while its sensitivity to $Q$ is substantially reduced. Thus, $k_r$ and $R/Q$ can be extracted robustly for very high-$Q$ modes, whereas the fitted $Q$ may remain weakly constrained unless the simulated wake length is comparable to the modal damping length. This behavior is illustrated by the pillbox benchmark discussed above. Because of the idealized perfectly conducting boundaries, the cavity modes have extremely large quality factors; therefore, the truncated spectra allow accurate extraction of the resonant frequencies and normalized shunt impedances, but provide limited sensitivity to $Q$.

For realistic RF cavities, however, wall losses, input couplers, waveguides, and dedicated HOM absorbers introduce additional damping, leading to substantially lower quality factors, particularly for higher-order modes. Under these conditions, the dependence of the truncated impedance spectrum on $Q$ becomes increasingly significant, allowing the quality factor to be extracted together with the resonant frequency and normalized shunt impedance.

The framework is next applied to the SRCLS RF cavity, a realistic three-dimensional structure incorporating HOM damping components. In addition to determining $f_r$, $R_{\parallel0}^{\mu}/Q$, and $R_{ud}^{\mu}/Q$, the analysis demonstrates the extraction of modal quality factors and their subsequent use in coupled-bunch-instability evaluation and cavity optimization.

\section{\label{sec:application}APPLICATION TO A 3D CAVITY}

Having validated the theoretical formulation and fitting procedure using the pillbox benchmark, we now apply the proposed framework to a practical RF cavity developed for the SRCLS storage ring. The objective is twofold: first, to extract the longitudinal and transverse resonant impedances of the cavity, with particular emphasis on higher-order modes (HOMs); second, to evaluate the corresponding beam-instability thresholds using the extracted modal parameters. Since HOMs are a major source of coupled-bunch instabilities in storage rings, such analyses provide quantitative guidance for cavity optimization and RF-system design~\cite{zheng2020higher,tavares2022beam,Bassi2016Analysis}. The principal machine and beam parameters of the SRCLS storage ring are summarized in Table~\ref{tab:table4_Key_par_SRCLS}.

\begin{table*}
\caption{\label{tab:table4_Key_par_SRCLS}Key machine and beam parameters of the SRCLS storage ring \cite{jiang2022synchrotron,LuYujie2025Lattice}.}
\begin{ruledtabular}
\begin{tabular}{lcl}%
Parameter & Symbols & Values \\%
\hline%
Beam energy & $E_0$ & $1400~\text{MeV}$ \\
Circumference & $C_0$ & $172.8~\text{m}$ \\%
Radiation loss per turn & $U_0$ & $1.10~\text{MeV}$ \\
Bunch length & $\sigma_z$ & $2.89~\text{mm}\ (\text{w/ IBS})$ \\
Harmonic number & $h$ & $288$ \\
Slip factor & $\eta$ & $7.177 \times 10^{-4}$ \\%
Average circulating current & $I$ & $1000~\text{mA}$ \\
Damping time (x,y,z) & $\tau_x/\tau_y/\tau_s$ & $1.46/1.46/0.73~\text{ms}$ \\
Tune & $\nu_s/\nu_x/\nu_y$ & $0.00956/19.14/9.23$ \\%
Average vertical beta function at the RF cavity & $\beta_{y,{cav}}$ & $4.40~\text{m}$ \\%
\end{tabular}
\end{ruledtabular}
\end{table*}

\begin{figure}[!h]
\includegraphics[width=6cm]{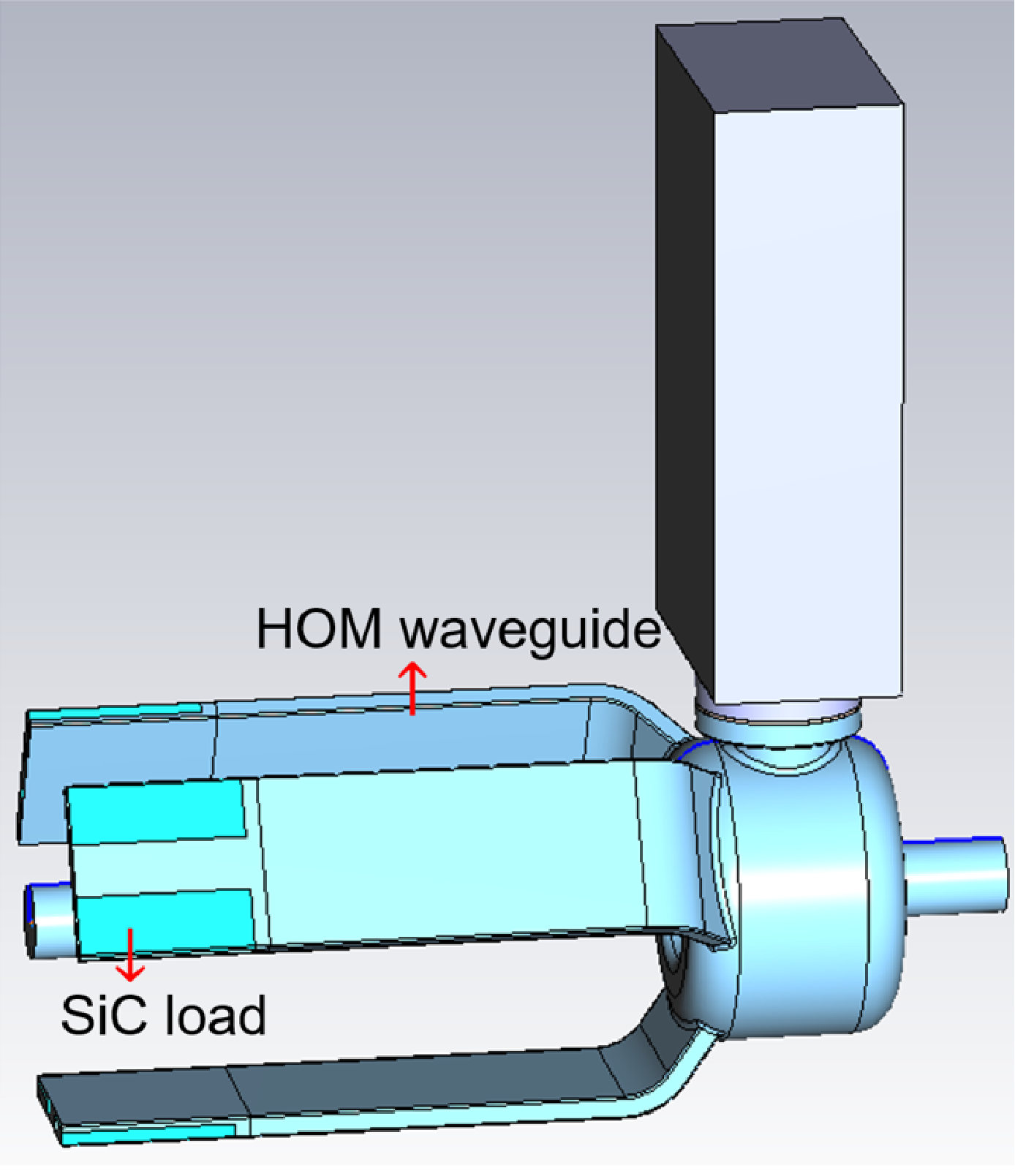}
\caption{\label{fig:fig_09_3Dcavity}Geometric model of the RF cavity of the SRCLS storage ring.}
\end{figure}

The RF cavity considered in this study is a normal-conducting cavity designed for the SRCLS storage ring. Its geometric model is shown in Fig.~\ref{fig:fig_09_3Dcavity}~\cite{Wang2025Design}. The cavity incorporates three HOM absorbers for suppressing parasitic resonances and an input coupler for RF power delivery. To compensate for the synchrotron-radiation energy loss per turn, a total of eight such cavities are required in the storage ring.

\subsection{\label{sec:LongitudinalImpedance}Longitudinal resonant impedance and instability threshold}

The choice of truncation lengths follows the analysis presented for the pillbox cavity and is therefore based on the modal damping coefficient $\alpha = k_r/(2Q)$ in Eqs.~\eqref{eq:Zz0FitModel} and~\eqref{eq:ZtdFitModel}. To ensure sufficient separation between the two truncation lengths for stable evaluation of the impedance difference and clear resolution of the resonant signatures, a pair of well-separated values is adopted. Following the truncation criteria established in the theoretical analysis and considering the computational cost of full three-dimensional wakefield simulations, we adopt $L_1 = 300~\mathrm{m}$ and $L_2 = 900~\mathrm{m}$ as representative truncation lengths.

\begin{figure}
\includegraphics[width=8cm]{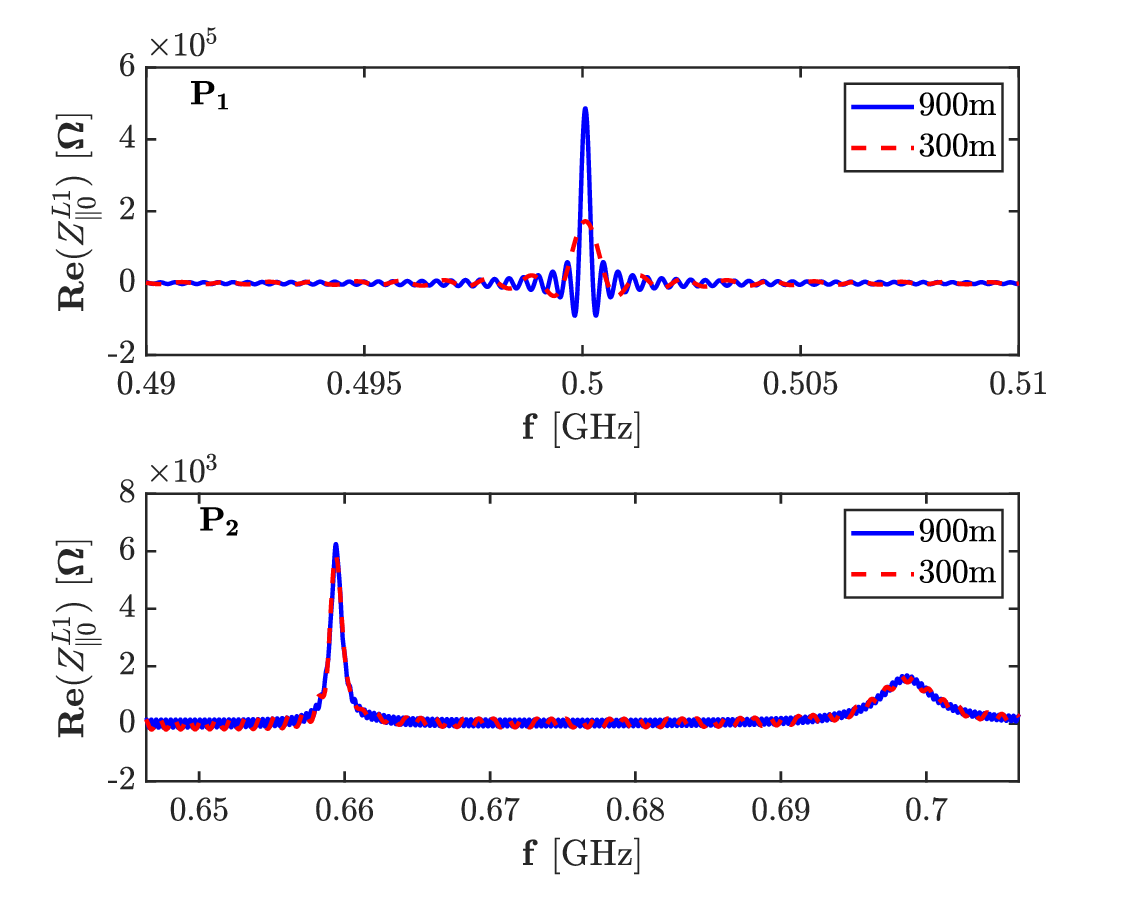}
\caption{\label{fig:fig_10_L_truncted_impedance_3D}Real part of the longitudinal impedance obtained from wake potentials truncated at $L_1=300$ m and $L_2=900$ m in the 3D cavity for the SRCLS RF cavity. Top: zoomed-in view around $0.500~\mathrm{GHz}$, corresponding to the fundamental mode. Bottom: zoomed-in view around $0.680~\mathrm{GHz}$.}
\end{figure}

Figure~\ref{fig:fig_10_L_truncted_impedance_3D} shows representative regions of the real part of the longitudinal impedance spectrum. As discussed in Sec.~\ref{sec:theory}, finite wakefield truncation produces oscillatory modulations around resonance peaks through the truncation factor in Eq.~\eqref{eq:TLsimple}. For high-$Q$ modes, the damping coefficient $\alpha$ is small, so the truncation effect remains visible over a wide frequency range; increasing $s_1$ mainly increases the modulation frequency, as illustrated by the fundamental resonance near $0.5~\mathrm{GHz}$.

For relatively low-$Q$ modes, however, $\exp(-\alpha s_1)$ can become small for sufficiently large $s_1$, causing $\tilde{T}_{\parallel}(k,s_1)$ to approach unity and weakening the truncation-induced modulation. Such modes are therefore less well constrained by the difference between two long truncation lengths. This behavior is observed for the resonance near $0.699~\mathrm{GHz}$, for which fitting with Eq.~\eqref{eq:Zz0FitModel} may be ineffective. In this case, the truncated impedance can be approximated by the standard resonator model in Eq.~\eqref{eq:longitudinal impedance}, effectively corresponding to a converged result. Alternatively, shorter truncation lengths may be used, although this option is not pursued here. The resonance near $0.659~\mathrm{GHz}$ represents an intermediate case, where the truncation-induced modulation remains sufficiently pronounced for fitting with Eq.~\eqref{eq:Zz0FitModel}.

Impedance peaks above $2~\mathrm{GHz}$ were excluded from the fitting because of their comparatively small amplitudes. The difference spectrum between the $300~\mathrm{m}$ and $900~\mathrm{m}$ calculations was then fitted using Eq.~\eqref{eq:Zz0FitModel}. Finally, the parameters of the resonant modes obtained from the fitting procedure are compared with those extracted from the eigenmode solver, as summarized in Table~\ref{tab:table5_01_3D_resonant_L}. Excellent agreement is observed between the two methods.

\begin{table*}
\caption{\label{tab:table5_01_3D_resonant_L}Comparison of SRCLS RF cavity resonant parameters obtained from truncated impedance fitting and eigenmode calculations for longitudinal resonant modes.}
\begin{ruledtabular}
\begin{tabular}{cccccc}%
\multicolumn{3}{c}{Eigenmode calculations} & \multicolumn{3}{c}{Truncated impedance fitting} \\
\hline
Frequency [\text{GHz}] &Q-value & $R_{\parallel 0}^{\mu}/{Q}$  &Frequency [\text{GHz}] & Q-value & $R_{\parallel 0}^{\mu}/{Q}$   \\
\hline
0.500   &30062     & 112.77  & 0.500       & 25973  & 112.83      \\
0.631   &41723     & 0.05    & 0.631       & 41179  & 0.05        \\
0.659   &802       & 8.24    & 0.658       & 743    & 9.38        \\
0.765   &2202      & 0.30    & 0.765       & 2091   & 0.19        \\
\end{tabular}
\end{ruledtabular}
\end{table*}

The threshold for longitudinal coupled-bunch instability is evaluated following Refs.~\cite{zheng2020higher,Wiedemann2007Particle,MarhauserF2001HOM},
\begin{eqnarray}
Z_{\parallel 0}^{th} = \frac{2 \nu_s \left( \frac{E_0}{e} \right)}{I_b f \eta \tau_s},
\label{eq:Rth_parallel}
\end{eqnarray}
where $\nu_s$ denotes the synchrotron tune, $E_0$ the beam energy, $e$ the elementary charge, $I_b$ the bunch current, $\eta$ the slip factor, and $\tau_s$ the longitudinal damping time.

\begin{figure}
\includegraphics[width=8cm]{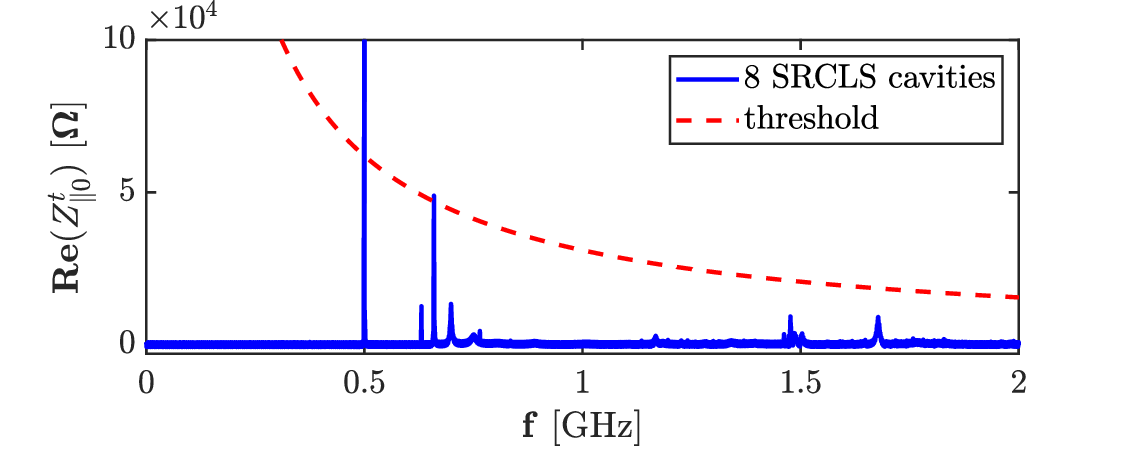}
\caption{\label{fig:fig_11_L_THandHOMs}Longitudinal total impedance spectra of the eight SRCLS cavities and the corresponding instability threshold.}
\end{figure}

The total longitudinal impedance is reconstructed by decomposing it into high-$Q$ contributions and the remaining impedance.
For high-$Q$ resonances such as the mode near $0.5~\mathrm{GHz}$, the corresponding impedance contributions are obtained using the modal parameters listed in Table~\ref{tab:table5_01_3D_resonant_L} together with Eq.~\eqref{eq:Zz0_single}. These same modal parameters are then applied in Eq.~\eqref{eq:Zz0truncation} to reconstruct the high-$Q$ truncated resonant impedance at each truncation length.
The remaining impedance is obtained by subtracting the reconstructed high-$Q$ truncated resonant impedance from the raw truncated impedance spectra computed at $L_1 = 300~\mathrm{m}$ or $L_2 = 900~\mathrm{m}$. This residual signal consists mainly of two distinct components: (i) low-$Q$ resonant modes not captured by the fitting procedure, and (ii) short-range wakefield contributions associated with fully damped modes. In addition, some weakly resonant modes with relatively high $Q$ but very small normalized shunt impedance $R/Q$ may also remain in the residual. These modes are not resolved by the fitting procedure and are neglected due to their negligible impact on beam dynamics and collective instabilities.
The high-$Q$ contributions and the remaining impedance are finally combined to form the complete longitudinal impedance spectrum.
The resulting complete impedance spectrum, together with the corresponding instability threshold, is shown in Fig.~\ref{fig:fig_11_L_THandHOMs}.

As shown in Fig.~\ref{fig:fig_11_L_THandHOMs}, the mode at $0.659~\mathrm{GHz}$ lies slightly above the instability threshold; however, the resulting growth rate is expected to remain modest and can be effectively suppressed by the longitudinal feedback system.

\subsection{\label{sec:VerticalImpedance}Vertical resonant impedance and instability threshold}

\begin{figure}
\includegraphics[width=8cm]{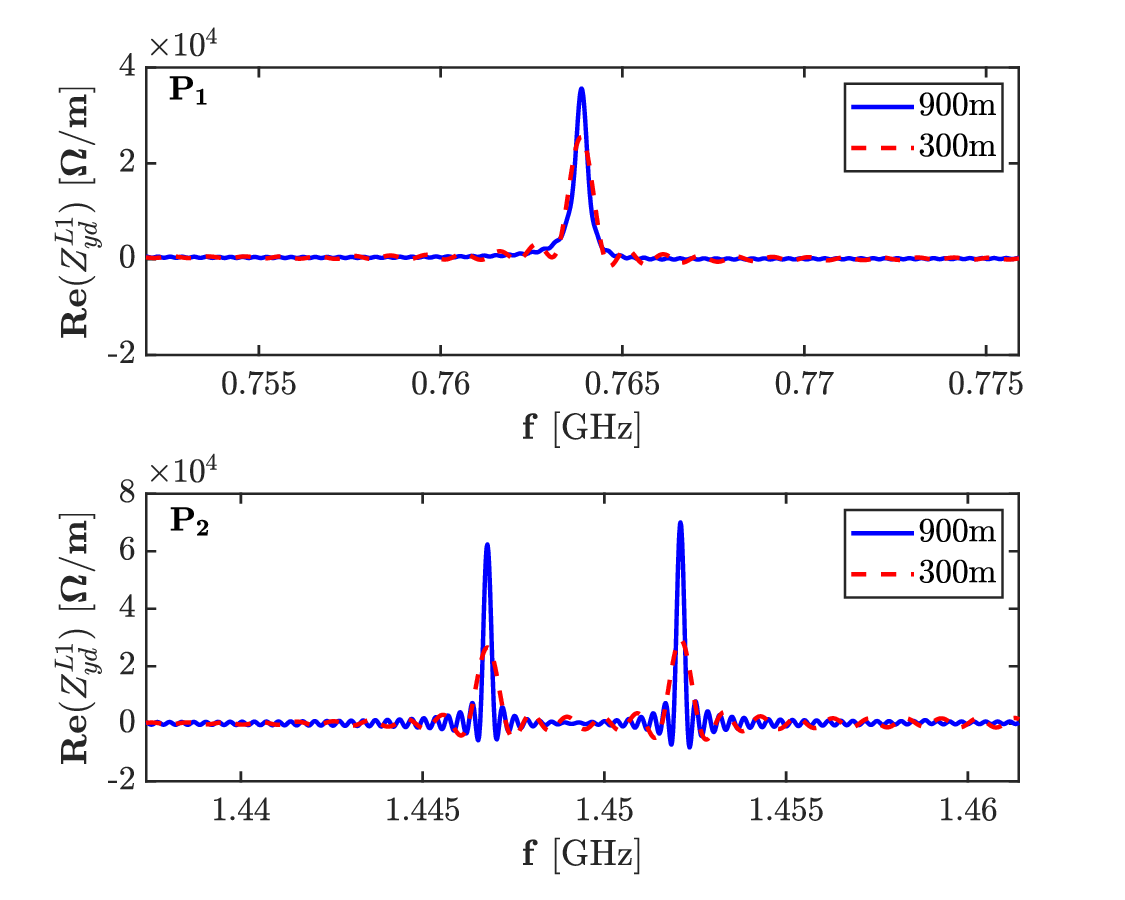}
\caption{\label{fig:fig_13_T_truncted_impedance_3D}Real part of the vertical dipolar impedance obtained from wake potentials truncated at $L_1=300$ m and $L_2=900$ m in the 3D cavity for the SRCLS RF cavity. Top: zoomed-in view around $0.765~\mathrm{GHz}$, corresponding to the fundamental mode. Bottom: zoomed-in view around $1.450~\mathrm{GHz}$.}
\end{figure}

To ensure sufficient excitation of the vertical dipolar wakefields while remaining within the linear field dynamics regime, symmetric beam offsets of $\pm 1~\mathrm{mm}$ were applied. The vertical dipolar impedance of the SRCLS cavity was then computed from wakefield simulations using truncation lengths $L_1=300~\mathrm{m}$ and $L_2=900~\mathrm{m}$, following the same choice as in the longitudinal impedance analysis. Representative portions of the real part of the resulting vertical impedance spectrum are shown in Fig.~\ref{fig:fig_13_T_truncted_impedance_3D}.

As in the longitudinal case, finite wakefield truncation produces oscillatory modulations around the resonance peaks through the truncation factor. This behavior is illustrated by the resonances shown in Fig.~\ref{fig:fig_13_T_truncted_impedance_3D}.
The mode near $0.765~\mathrm{GHz}$, shown in subplot $P_1$, is analogous to the intermediate-$Q$ case observed for the longitudinal resonance near $0.659~\mathrm{GHz}$.
In contrast, the modes near $1.448~\mathrm{GHz}$ and $1.454~\mathrm{GHz}$, shown in subplot $P_2$, exhibit behavior similar to the high-$Q$ longitudinal resonance near $0.5~\mathrm{GHz}$. In this regime, increasing $s_1$ mainly increases the modulation frequency in the impedance spectrum, while the oscillatory structure remains pronounced.

Impedance peaks above $2~\mathrm{GHz}$ are excluded from the fitting procedure because their amplitudes are comparatively small. The resonant-mode parameters obtained from the fitting are then compared with those extracted from the eigenmode solver, as summarized in Table~\ref{tab:table6_01_3D_th_T}. As in the longitudinal case, excellent agreement is observed in the vertical dipolar modes between the fitted parameters and those obtained from eigenmode calculations. For comparison, the table also lists the dipolar shunt impedances evaluated from the eigenmode results using both Eqs.~\eqref{eq:cal_Rs_eigenmode_d} and~\eqref{eq:cal_Rs_eigenmode_d_arsenyev}. In practice, the CST Eigenmode Solver can directly output the shunt impedance $\mathcal{R}_\parallel$ defined in Eq.~\eqref{eq:RLeigenmode}. From this output, one can compute
$R_{ud}^{AS}$ according to Eq.~\eqref{eq:cal_Rs_eigenmode_d_arsenyev}. By contrast, the evaluation of $R_{ud}$ defined by Eq.~\eqref{eq:cal_Rs_eigenmode_d} requires additional post-processing. Specifically, the three-dimensional field data must first be exported from the CST Eigenmode Solver, after which the complex voltage $\mathcal{V}_0(\mathbf{r}_{0\perp})$ is obtained by numerical integration along a predefined path. The required transverse derivatives can then be evaluated from the resulting offset-dependent complex voltage, yielding the corresponding value of $R_{ud}$.

\begin{table*}
\caption{\label{tab:table6_01_3D_th_T}Comparison of SRCLS RF cavity resonant parameters obtained from truncated impedance fitting and eigenmode calculations for vertical resonant modes.} 
\begin{ruledtabular}
\begin{tabular}{ccccccc}%
\multicolumn{4}{c}{Eigenmode calculations} & \multicolumn{3}{c}{Truncated impedance fitting} \\
\hline
Frequency  &Q-value & $R_{u d}^{\mu}/{Q }$ (Eq.~\eqref{eq:cal_Rs_eigenmode_d}) & $R_{u d}^{\mu}/{Q }$ (Eq.~\eqref{eq:cal_Rs_eigenmode_d_arsenyev}~\cite{arsenyev2019method}) &Frequency  & Q-value & $R_{u d}^{\mu}/{Q }$   \\
$[\text{GHz}]$&&$[{\Omega/\text{m}^2}]$&$[{\Omega/\text{m}^2}]$&$[\text{GHz}]$&&$[{\Omega/\text{m}^2}]$\\
\hline
0.765   &2202     & 278  & 277     &0.765   & 2088   & 292       \\
1.423  &3581     & 323  & 334     &1.422  & 3704   & 316       \\
1.448  &17054    & 190  & 185     &1.448  & 16729  & 190       \\
1.454  &22635    & 175  & 186     &1.453  & 21425  & 182       \\
1.473  &5372     & 944  & 1194    &1.473  & 4780   & 1013      \\
1.592  &5549     & 96   & 82      &1.591  & 5734   & 105       \\
\end{tabular}
\end{ruledtabular}
\end{table*}

Overall, the dipolar shunt impedances obtained using Eq.~\eqref{eq:cal_Rs_eigenmode_d} agree better with the truncated-impedance fitting results for most of the identified HOMs. This suggests that the complex-voltage-derivative formulation consistently reproduces the impedance parameters derived from the truncated spectral representation. An exception occurs for the mode near $1.453~\mathrm{GHz}$, for which the value obtained using Eq.~\eqref{eq:cal_Rs_eigenmode_d_arsenyev} is slightly closer to the truncated-fitting result. A detailed analysis shows that the field profile of this mode is nearly symmetric with respect to the center position, and that the field amplitude at the center is close to zero. Following Ref.~\cite{arsenyev2019method}, the zero-crossing, or tangent-constrained, fitting procedure is therefore applied. However, this constraint does not account for the non-negligible quadrupolar contribution near the field zero. Consequently, part of the quadrupolar variation is interpreted as a dipolar component, leading to a slight overestimation of the extracted dipolar impedance. The resulting error partially compensates for the intrinsic difference between the two extraction methods, so that the final value appears closer to that obtained from the truncated-impedance fitting.

The threshold transverse impedance associated with vertical coupled-bunch instability is given by~\cite{zheng2020higher,chao2023handbook,MarhauserF2001HOM}
\begin{eqnarray}
Z_{u d}^{th} = \frac{2 {E_0}/{e} }{f_{\text{rev}} I_b \beta_{y,cav} \tau_y},
\label{eq:Rx_th}
\end{eqnarray}
where $f_{\mathrm{rev}}$ is the revolution frequency, $\beta_{y,\mathrm{cav}}$ is the average vertical beta function at the RF cavity, and $\tau_y$ is the vertical damping time.

\begin{figure}
\includegraphics[width=8cm]{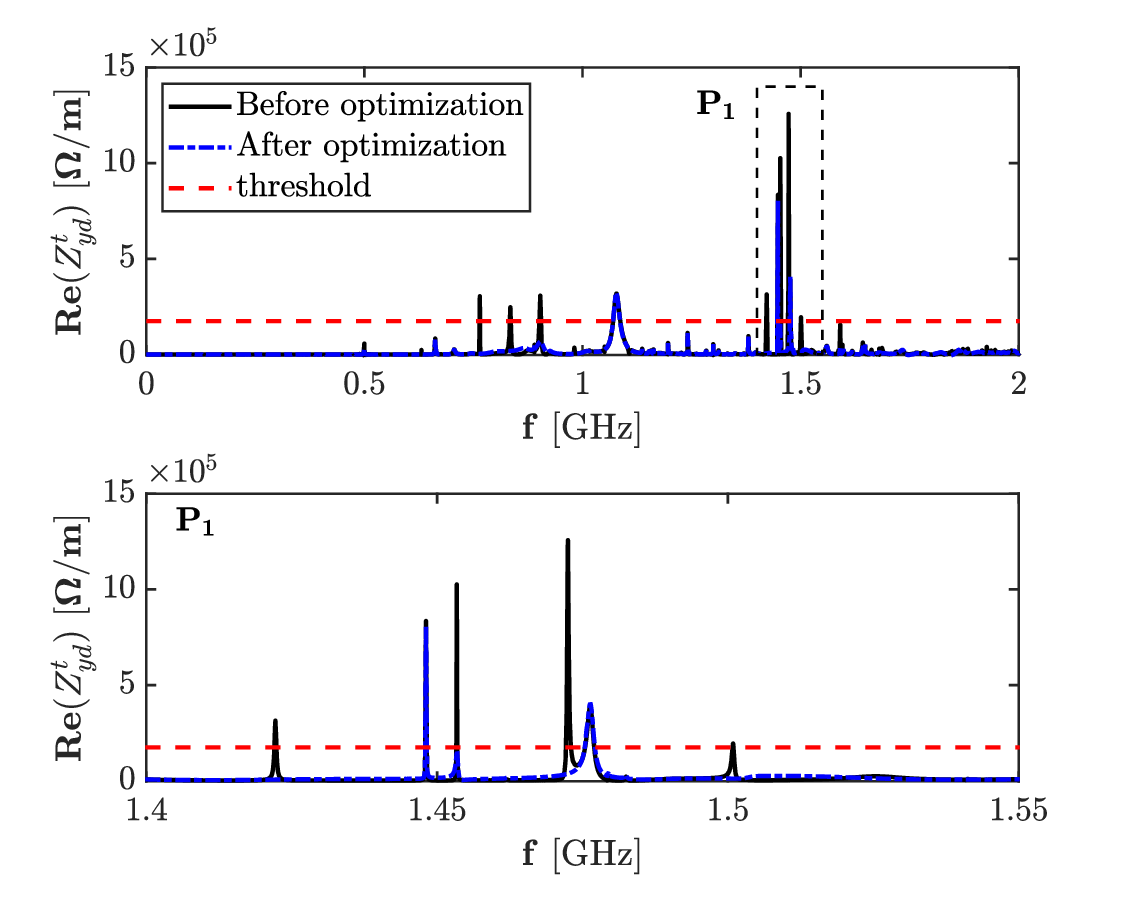}
\caption{\label{fig:fig_17_V_THandHOMs_boths}Comparison of the vertical total impedance spectra of the eight SRCLS cavities before and after geometry optimization, together with the corresponding instability threshold. Top: full spectrum up to $2~\mathrm{GHz}$. Bottom: zoomed-in view around $1.453~\mathrm{GHz}$.}
\end{figure}

Following the same procedure as for the longitudinal impedance, the total transverse impedance spectrum is reconstructed by combining the fitted parameters of the high-$Q$ transverse resonant modes listed in Table~\ref{tab:table6_01_3D_th_T} with a decomposition of the truncated wakefield data.
The high-$Q$ contributions are reconstructed using Eq.~\eqref{eq:Ztd} and the corresponding modal parameters from Table~\ref{tab:table6_01_3D_th_T}, and their truncated counterparts are obtained using Eq.~\eqref{eq:Ztdtruncation}.
The resulting total impedance spectrum is shown by the black curves in Fig.~\ref{fig:fig_17_V_THandHOMs_boths}, together with the instability threshold defined by Eq.~\eqref{eq:Rx_th}.

As indicated by the black curves in Fig.~\ref{fig:fig_17_V_THandHOMs_boths}, several impedance peaks in the original cavity exceed the instability threshold. In particular, the modes at $1.448~\mathrm{GHz}$, $1.453~\mathrm{GHz}$, and $1.473~\mathrm{GHz}$ lie well above the threshold, indicating that they are likely to dominate the vertical coupled-bunch instability. Further optimization of the cavity geometry is therefore required to reduce the impedances of these resonant modes.

\begin{figure}[!h]
\includegraphics[width=6cm]{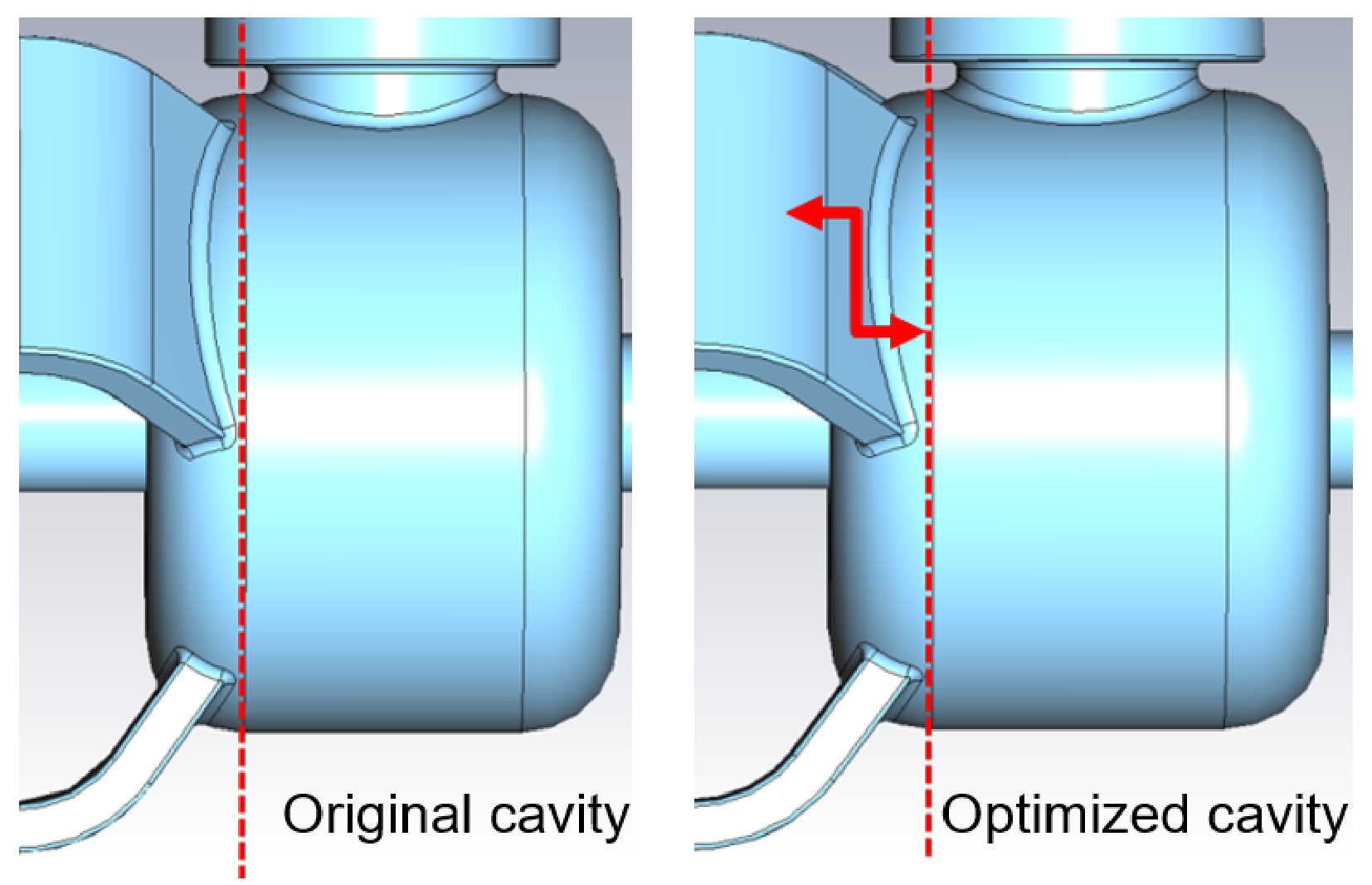}
\caption{\label{fig:fig_16_01_3Dcavity}Schematic illustration of the longitudinal offset arrangement of the HOM waveguides.}
\end{figure}

To mitigate the vertical coupled-bunch instability identified above, the cavity geometry was optimized by shifting the two upper HOM waveguides longitudinally in opposite directions by $5~\mathrm{mm}$ from their original positions at the cavity junctions, as illustrated in Fig.~\ref{fig:fig_16_01_3Dcavity}. This modification breaks the rotational symmetry of the HOM-coupler geometry and is intended to suppress the excitation of the dominant vertical dipolar modes.

Using the same fitting procedure described previously, the resonant frequencies, normalized dipolar shunt impedances $R_{ud}^{\mu}/Q$, and the corresponding quality factors were re-evaluated for the optimized cavity. The resulting resonant-mode parameters are summarized in Table~\ref{tab:table6_3D_th_T_optimize}. Using these parameters, the total vertical impedance spectrum of the optimized cavity was reconstructed. The optimized spectrum is shown by the blue curves in Fig.~\ref{fig:fig_17_V_THandHOMs_boths}, enabling a direct comparison with the original design.
\begin{table}
\caption{\label{tab:table6_3D_th_T_optimize}Fitted modal parameters for the optimized SRCLS RF cavity.}
\begin{ruledtabular}
\begin{tabular}{ccc}%
Frequency  & Q-value & $R_{u d}^{\mu}/{Q }$ \\%
$[\text{GHz}]$  &  & $[{\Omega/\text{m}^2}]$ \\%
\hline%
1.241   & 7908     & 45  \\
1.380   & 1821     & 186   \\
1.448   & 16090    & 187 \\
1.453   & 3625     & 163 \\
1.476   & 4756     & 35   \\
\end{tabular}
\end{ruledtabular}
\end{table}

Comparison of the modal parameters in Tables~\ref{tab:table6_01_3D_th_T} and~\ref{tab:table6_3D_th_T_optimize}, together with the impedance spectra shown in Fig.~\ref{fig:fig_17_V_THandHOMs_boths}, demonstrates that the dominant vertical dipolar resonances are significantly suppressed after the introduction of the waveguide offset. In particular, the most critical mode near $1.473~\mathrm{GHz}$ exhibits a substantial reduction in $R_{ud}^{\mu}/Q$, leading to a marked decrease in the corresponding impedance peak. These results demonstrate that a relatively simple geometric modification can substantially reduce the dominant transverse HOM impedances and thereby mitigate the risk of coupled-bunch instability.

Nevertheless, impedance peaks near $1.079~\mathrm{GHz}$, $1.448~\mathrm{GHz}$, and $1.476~\mathrm{GHz}$ remain above the instability threshold after optimization. As shown in Table~\ref{tab:table6_3D_th_T_optimize}, the $1.448~\mathrm{GHz}$ mode still possesses both a relatively large quality factor and a high value of $R_{ud}^{\mu}/Q$, resulting in a significant contribution to the transverse impedance. In contrast, the mode near $1.476~\mathrm{GHz}$ is characterized by moderate values of both $Q$ and $R_{ud}^{\mu}/Q$, but overlaps with a neighboring low-$Q$ resonance having a comparatively large $R_{ud}^{\mu}/Q$, giving rise to an elevated combined impedance peak. The peak near $1.079~\mathrm{GHz}$ is likewise associated with a low-$Q$ resonance possessing a relatively large dipolar shunt impedance.

Although some remaining peaks still exceed the instability threshold, their amplitudes are substantially reduced compared with those before optimization. The associated instability growth rates are therefore expected to be substantially reduced and may remain within the suppression capability of a modern transverse feedback system. A quantitative evaluation of the coupled-bunch growth rates and feedback damping rates is left for future work. It should also be noted that the present study is intended as a proof-of-principle example of HOM-oriented cavity-geometry optimization rather than as a complete engineering optimization. More systematic strategies, including automated algorithms for determining the absorber-waveguide position and tilt angle, are expected to further reduce the dominant HOM impedances and improve the transverse beam-stability margin.

\section{\label{sec:summary}CONCLUSIONS}

In this work, we developed a unified framework for describing longitudinal and transverse resonant impedances of RF cavities based on Maxwell's equations and generalized cavity-voltage definitions. The formulation connects wakefield-based impedance definitions with eigenmode quantities and provides consistent expressions for monopolar, dipolar, and quadrupolar resonant impedance components in both symmetric and three-dimensional asymmetric cavities.

Analytical expressions were derived for impedances obtained from finite-length truncated wakefields, allowing resonant frequencies, normalized shunt impedances, and, when sufficiently constrained, quality factors to be extracted without fully converged long-range wake simulations. The method was validated with an axisymmetric pillbox cavity, showing excellent agreement with analytical results and CST eigenmode calculations.

The framework was then applied to the SRCLS RF cavity. Longitudinal monopolar and vertical dipolar HOM parameters were extracted and used to reconstruct the total impedance spectra of the eight cavities and evaluate coupled-bunch instability thresholds. The analysis identified several vertical HOMs exceeding the threshold, and a simple HOM-waveguide offset was shown to substantially reduce the dominant transverse impedance peaks. These results demonstrate that the proposed method provides an efficient and physically transparent link between finite-length wakefield simulations, eigenmode characterization, and beam-stability evaluation, making it useful for HOM identification and RF-cavity optimization in high-current storage rings.

\section*{Acknowledgments}
This work was supported by the Shanghai Municipal Science and Technology Major Project, the Special Fund Project for Improving Scientific Research Conditions, and Zhangjiang Laboratory. We also thank Dr. Dinghui Su and Wencheng Fang (SARI) for useful discussions on the optimization of RF cavity structures.

\appendix
\section{\label{sec:WakeFunctionApprox} Approximate longitudinal wake function for high-$Q$ modes}

The exact solution of Eq.~\eqref{eq:Wake_function} can be expressed as
\begin{equation}
    W_\parallel(s)=
    \frac{1}{U_0} \left[
        F_c'\left({s}\right) V^+
        +F_s'\left({s} \right) V^-
    \right],
    \label{eq:wakeExact}
\end{equation}
where
\begin{equation}
    F_s'(x)=e^{-\alpha x}
    \left[
    \sin\!\bigl(k_d x\bigr)
    +\frac{\alpha}{k_d}\cos\!\bigl(k_d x \bigr)
    \right],
\end{equation}
\begin{equation}
    V^+=\frac{1}{2} \left[ V_1^*(\mathbf r_{1\perp})V_0(\mathbf r_{0\perp}) + V_1(\mathbf r_{1\perp})V_0^*(\mathbf r_{0\perp})  \right],
\end{equation}
\begin{equation}
    V^-=\frac{i}{2} \left[ V_1^*(\mathbf r_{1\perp})V_0(\mathbf r_{0\perp}) - V_1(\mathbf r_{1\perp})V_0^*(\mathbf r_{0\perp})  \right],
\end{equation}
with
\begin{equation}
V_0(\mathbf{r}_{0\perp}) = \int_{0}^{L_c} e^{-(\gamma + i\omega_d)\frac{z}{c}} E_{0z}(x_0, y_0, z) dz
\end{equation}
and
\begin{equation}
V_1(\mathbf{r}_{1\perp}) = \int_{0}^{L_c} e^{-(\gamma + i\omega_d)\frac{z}{c}} E_{0z}(x_1, y_1, z) dz.
\end{equation}

For high-$Q$ modes, $\gamma\ll\omega_d$, and the damping of the fields during the passage of the source and test particles through the cavity is negligible. $V_0$ and $V_1$ can be respectively approximated by
\begin{equation}
\mathcal{V}_0(\mathbf{r}_{0\perp}) = \int_{0}^{L_c} e^{- i\omega_r\frac{z}{c}} E_{0z}(x_0, y_0, z) dz
\label{eq:AccVoltageSource}
\end{equation}
and
\begin{equation}
\mathcal{V}_1(\mathbf{r}_{1\perp}) = \int_{0}^{L_c} e^{-i\omega_r\frac{z}{c}} E_{0z}(x_1, y_1, z) dz.
\label{eq:AccVoltageTest}
\end{equation}

For small transverse offsets of the source and test particles, the two terms inside the square brackets of $V^-$ are nearly equal, so that $V^-$ is small compared with $V^+$. In this case, Eq.~\eqref{eq:wakeExact} reduces to
\begin{equation}
    W_\parallel(s) =
    \frac{1}{U_0}
        F_c'\left({s}\right) \mathcal{V}^+,
    \label{eq:wakeApproximate}
\end{equation}
where
\begin{equation}
    \mathcal{V}^+=\frac{1}{2} \left[ \mathcal{V}_1^*(\mathbf r_{1\perp})\mathcal{V}_0(\mathbf r_{0\perp}) + \mathcal{V}_1(\mathbf r_{1\perp})\mathcal{V}_0^*(\mathbf r_{0\perp})  \right].
    \label{eq:effectiveVoltage}
\end{equation}

\bibliography{paper_wang_01}

\end{document}